\documentclass[12pt]{article}
\usepackage{amsmath}
\usepackage{graphicx}
\usepackage{enumerate}
\usepackage{natbib}
\usepackage{url} 

\usepackage{blindtext}
\usepackage{hyperref}
\usepackage[capitalise, noabbrev, english]{cleveref}
\crefname{equation}{}{}
\crefname{appendix}{Supplementary material}{Supplementary material}

\usepackage{mathabx}
\usepackage{multirow}
\usepackage{caption}
\usepackage{subcaption}

\usepackage{amsmath}

\usepackage{graphicx}
\usepackage{enumerate}

\usepackage{lscape}

 \usepackage{tikz}
 \usetikzlibrary{shapes,arrows}

\usepackage{placeins}

\newcommand{\blind}{1}

\begin{document}

\def\spacingset#1{\renewcommand{\baselinestretch}%
{#1}\small\normalsize} \spacingset{1}


\if1\blind
{
  \title{\bf A Shared Parameter Model for Systolic Blood Pressure Accounting for Data Missing Not at Random in the HUNT Study.}
  \author{Aurora Christine Hofman
    \hspace{.2cm}\\
    Department of Mathematical Sciences\\
    NTNU (Norwegian University of Science and Technology)\\
    and \\
    Lars Espeland \\
    Department of Mathematical Sciences \\
    NTNU (Norwegian University of Science and Technology)\\
        and \\
    Ingelin Steinsland \\
   Department of Mathematical Sciences\\
   NTNU (Norwegian University of Science and Technology)\\
            and \\
    Emma M. L. Ingeström \\
    Department of Circulation and Medical Imaging\\
    NTNU (Norwegian University of Science and Technology)}
  \maketitle
} \fi

\if0\blind
{
  \bigskip
  \bigskip
  \bigskip
  \begin{center}
    {\LARGE\bf A Shared Parameter Model for Systolic Blood Pressure Accounting for Data Missing Not at Random in the HUNT Study.}
\end{center}
  \medskip
} \fi

\bigskip
\begin{abstract}

In this work, blood pressure eleven years ahead is modeled using data from a longitudinal population-based health survey, the Trøndelag Health (HUNT) Study, while accounting for missing data due to dropout between consecutive surveys ($20-50 \%$). We propose and validate a shared parameter model (SPM) in the Bayesian framework with age, sex, body mass index, and initial blood pressure as explanatory variables. Further, we propose a novel evaluation scheme to assess data missing not at random (MNAR) by comparing the predictive performance of the fitted SPM with and without conditioning on the missing process. The results demonstrate that the SPM is suitable for inference for a dataset of this size (cohort of $64385$ participants) and structure. The SPM indicates data MNAR and gives different parameter estimates than a naive model assuming data missing at random. The SPM and naive models are compared based on predictive performance in a validation dataset. The naive model performs slightly better than the SPM for the present participants. This is in accordance with results from a simulation study based on the SPM where we find that the naive model performs better for the present participants, while the SPM performs better for the dropouts.

\end{abstract}

\noindent%
{\it Keywords:} 
longitudinal studies, missing data,  INLA (integrated nested Laplace approximations), dropout, health survey
\vfill     

\newpage 
\spacingset{1.9} 
\section{Introduction}
\label{sec:intro}

This work aims to establish and validate a predictive model for future systolic blood pressure using data from a longitudinal population-based health survey, the Trøndelag Health (HUNT) Study, while accounting for missing data due to dropout. 

Elevated blood pressure increases the risk of developing diseases related to the brain, heart, blood vessels, and kidney \citep{prospective2002age, tozawa2003blood, rapsomaniki2014blood}. It affects more than 1.1 billion people worldwide and accounts for over 10.8 million deaths per year, thereby surpassing smoking as the leading preventable cause of death for middle-aged and older adults \citep{zhou2017worldwide, murray2020global}. Early detection, prevention, and treatment of elevated blood pressure are of high priority in public health strategies \citep{who_hypertention}. Thus, obtaining unbiased, accurate models for predicting future blood pressure is of great interest in medical research \citep{whelton1994epidemiology}. Longitudinal population-based health survey provide valuable datasets for constructing such models but are rarely without missing data. For instance, each HUNT survey include 50000 to 80000 participants, but $20-50\%$ of the participants are lost to follow-up in consecutive surveys \citep{krokstad2013cohort, aasvold2021cohort}. Proper handling of missing data is vital to obtain unbiased inference \citep[Chap.~1.3, 6]{gad2013shared, little2019statistical}, and how to 
handle missing data depends on the missing process.

Based on available literature \citep{anderson1994effect, whelton1994epidemiology, brown2000body, jiang2016obesity, master_Lars}, we suggest a predictive model of future blood pressure 
with age, sex, 
body mass index (BMI), and initial blood pressure as explanatory variables. All participants used in this study have full records for the explanatory variables.
Missing data can be categorized and described in terms of three missing processes; missing completely at random (MCAR), missing at random (MAR), and missing not at random (MNAR) \citep[Chap.~1.3, 6]{little2019statistical}. 
If the probability of dropout, i.e. of missing 
future blood pressure values, is independent of all observed and unobserved data, including the missing response variables and all explanatory variables, the data is MCAR. It is reasonable that the probability of dropping out depends on age, and hence MCAR is disregarded for this work. The data is MAR if the probability of missingness depends on the observed data, but is independent of the unobserved data.
Missing processes that are MCAR or MAR are ignorable, meaning unbiased inference can be performed without modeling the missing process. 
Data that is neither MCAR nor MAR is MNAR \citep[Chap.~6]{gad2013shared, little2019statistical}. 
If the part of the future blood pressure 
which can not be explained by 
age, sex, BMI, and initial blood pressure,  affects the probability of dropping out, the data is MNAR. 
This can be thought of as (unknown) explanatory variables not included in the models.
It is reasonable to assume that there are health related variables that influence both blood pressure and the probability of drop out. 
Thus, we argue that in a predictive model for future blood pressure
we should consider that data might be MNAR.
If data is MNAR the missing process must be modeled simultaneously with the original model to obtain unbiased inference \citep[Chap.~1.3, 6]{little2019statistical}. 

Even though the assumption of data MAR is often not fulfilled, many of the available software packages and methods described in the literature assume data to be MAR \citep{ balakrishnan2009methods, rhoads2012problems, little2019statistical, mohan2021graphical, Grisewold2021}. However, several studies, especially in biostatistics, have accounted for missing data under the assumption of data MNAR. \citep{wu1988estimation, little1993pattern, diggle1994informative, follmann1995approximate,   little1995modeling, albert2000modeling, molenberghs2008every, howe2016selection}. Popular choices for models accounting for data MNAR include the pattern mixture model, selection model, and  shared parameter model (SPM) \citep[~Chap. 15.4]{heckman1979sample, wu1988estimation, little1993pattern, henderson2000joint, linero2018bayesian, little2019statistical, Grisewold2021}. The SPM is based on the idea of a commonly shared variable affecting both the measurement process and the missing process. Given this variable, the two marginal densities are conditionally independent. It has been used to model longitudinal data subject to MNAR in several studies
\citep{wu1988estimation, follmann1995approximate, thomas1998mixed, pulkstenis1998model, vonesh2006shared, creemers2010sensitivity}.
In this work, we propose a Bayesian SPM for future blood pressure. The model fits the framework of Bayesian latent Gaussian model
and is suitable for Bayesian inference using computationally efficient Integrated Nested Laplace Approximations (INLA) \citep{rue2009approximate,rue2017bayesian, martino2019integrated, gomez2020bayesian, steinsland2014quantitative}. 

\citeauthor{molenberghs2008every} stated that "each MNAR model fit to a set of observed data can be reproduced exactly by a MAR counterpart" \citep[p.~371]{molenberghs2008every}. Hence, the choice between models eventually comes down to choosing the most likely model assumptions \citep{enders2011missing}. Recent research has proven that taking the approach of causal modeling and formulating the models through missingness graphs can give theoretical understanding and asymptotic performance guarantees \citep{mohan2021graphical}.
To the best of our knowledge, the literature provides little insight into practical validation of model performance on data MNAR. The current standard seems to be the use of simulation studies to check the reproducibility of the model. i.e., how well the original parameters are reproduced on simulated data, and sensitivity analysis to check the robustness of the models \citep{enders2011missing, steinsland2014quantitative, kaciroti2021bayesian}. 

In this work, we validate the models on a validation dataset. First, predictive performance of the SPM and a naive model assuming the data to be MAR are compared based on the proper scoring rules  \citep{gneiting2007strictly} continuous ranked probability score (CRPS) and Brier score. Second, we propose a new method to evaluate if data is MNAR based on the SPM. The main idea is that if data is MNAR, the missing status has information about the quantity of interest. Therefore, we compare the predictive performance of the SPM with and without conditioning on missing status. 

The main contributions of this paper is the SPM for future blood pressure based on data from the HUNT Study, the demonstration of the SPM's applicability for a large case study and new insight from the proposed validation schemes.

\Cref{sec:bg} provides background about latent Gaussian models and missing data theory. \Cref{sec:m&m} introduces the blood pressure case study including the HUNT Study, the proposed models, and methods for inference and validation. The results from the case study are presented in \Cref{sec:results}. 
\Cref{sec:simulation_study} consists of several simulation sensitivity studies based on the HUNT Study. \cref{sec:discussion} summarizes and discusses our findings.

\section{Background Theory}
\label{sec:bg}

This section briefly introduces latent Gaussian models and commonly used models and methods for missing data.

\subsection{Latent Gaussian Models}
\label{sec:bg_latent_gaussion_models}
Latent Gaussian models (LGMs) fall within a subclass of the structured additive regression models \citep{rue2009approximate} meaning the response $y_i$ belongs to the class of exponential families. Hence, the mean $E(y_i) = \mu$ is linked to a structured additive predictor $\eta$ through a link function $h(\mu)$ such that $h(\mu) = \eta$. For the structured additive regression models $\eta$ is defined as follows \citep{fahrmeir2007regression}, 
\begin{equation*}
    h(\mu) = \eta = \alpha + \sum_{k = 1}^{n_\beta} \beta_k z_{k} + \sum_{j = 1}^{n_f}f^{(j)}(u_{j}) + \epsilon.
\end{equation*}
Here $\{ \beta_k \}$ represents the linear effects of explanatory variables $\boldsymbol{z}$, $\{ f^{(j)} (.)\}$ represents unknown functions of explanatory variables $\boldsymbol{u}$, and $\epsilon$ is an unstructured term. All LGMs have Gaussian prior distributions of  $\alpha$, $\{ \beta_k \}$, $\{ f^{(j)} (.)\}$ and $\epsilon$. 
All models used in this work belong to the class of LGMs. 

\subsection{Missing Data}
In this section we follow \cite{little2019statistical} and let $\boldsymbol{y_i}$ be the set of $j$ measurements on the $i$th subject. Then $\boldsymbol{y_i}$ can be divided into an observed part $\boldsymbol{{y_i}_o}$ and a missing part $\boldsymbol{{y_i}_m}$,  $\boldsymbol{y_i} = (\boldsymbol{{y_i}_o}, \boldsymbol{{y_i}_m})$.  Let $\boldsymbol{m_i}$ be the vector of,
\begin{equation*}
  m_{ij} =
    \begin{cases}
      1 & \text{if ${y_{ij}}$ is missing}\\
      0 & \text{otherwise}.
    \end{cases}       
\end{equation*}
Then the full conditional of $\boldsymbol{y_i}$ and $\boldsymbol{m_i}$ is given as follows, 
\begin{equation}
\label{eq:full_cond}
    g(\boldsymbol{{y_i}_o},\boldsymbol{{y_i}_m}, \boldsymbol{m_i}| \boldsymbol{\theta}, \boldsymbol{\psi}) 
\end{equation}
where the parameters $\boldsymbol{\theta}$ and $\boldsymbol{\psi}$ describes the measurement process and missing process, respectively \citep[Chap.~6.2]{gad2013shared, little2019statistical}.
The data is MCAR if the missing process 
    $g(\boldsymbol{m_i} |  \boldsymbol{{y_i}_o},\boldsymbol{{y_i}_m}, \boldsymbol{\psi}) = g(\boldsymbol{m_i} |  \boldsymbol{\psi})$.
The data is MAR if 
$g(\boldsymbol{m} |  \boldsymbol{{y_i}_o},\boldsymbol{{y_i}_m}, \boldsymbol{\psi}) = g(\boldsymbol{m_i} |\boldsymbol{{y_i}_o},  \boldsymbol{\psi})$.
If the data is neither MCAR nor MAR, the data is, by definition, MNAR. 
If data is MNAR, the missing process must be modeled simultaneously with the measurement process to obtain unbiased inference \citep[Chap.~6.2]{little2019statistical}, and 
several models have been proposed including pattern mixture models and selection models \citep{little2019statistical}.
In this work, a class of selection models known as shared parameter models (SPMs) is used.
From now on, let $\boldsymbol{x_i}$ be the set of fully observed explanatory variables and $\epsilon_i$ be an unobserved within-subject random effect with hyperparameter $\gamma$.
\citep[Chap.~15.2]{little2019statistical} defined SPM as follows:
\begin{equation}
\label{eq:spm_bg}
    g(\boldsymbol{y_i}, \boldsymbol{m_i} , \epsilon_i| \boldsymbol{x_i}, \boldsymbol{\theta}, \boldsymbol{\psi}, \gamma) = 
    g(\boldsymbol{y_i}|  \boldsymbol{x_i}, \epsilon, \boldsymbol{\theta}) 
    g(\boldsymbol{m_i}| \boldsymbol{x_i}, \epsilon_i, \boldsymbol{\psi})
    g(\epsilon_i | \boldsymbol{x_i}, \gamma).
\end{equation}
This model assumes that both the measurement and dropout processes depend on a shared latent variable $\epsilon_i$.
MAR is then a special case with $g(m_i| \boldsymbol{x_i}, \epsilon_i, \boldsymbol{\psi}) = g(m_i| \boldsymbol{x_i}, \boldsymbol{\psi})$ \citep{vonesh2006shared}. 

\section{Case Study: A Blood Pressure Predictive Model Based on the HUNT Study.}
\label{sec:m&m}

\subsection{The HUNT Study and Explanatory Analyses}
\label{sec:data}
The HUNT Study is a longitudinal population-based health survey in central Norway and the study protocols have been described in detail previously by \citep{krokstad2013cohort, aasvold2021cohort} (\cref{sec:hunt_protocol}).
Every adult citizen in the now former county of Nord-Trøndelag were invited to participate in  clinical examinations and questionnaires in 1984-86 (HUNT1), 1995-97 (HUNT2), 2006-08 (HUNT3), and 2017-19 (HUNT4) \citep{krokstad2013cohort, aasvold2021cohort}. 
In this study, observations of systolic blood pressure ($BP$), age ($age$), body mass index ($BMI$) and sex ($sex$, 0 for females and 1 for males) are used.
Following \cite{tobin2005adjusting} $BP$ is adjusted by adding 15 mmHg for all participants who self-reported using BP medication. When needed a subscript indicates the HUNT survey of the observation (e.g. $BP_2$ denotes BP observed at HUNT2). 
We define a training dataset (HUNT2 cohort) with observations of initial blood pressure
($BP_I=BP_2$), $age$, $BMI$ and $sex$ from HUNT2, together with future blood pressure ($BP_F=BP_3$) from HUNT3 and a missing indicator $m$ (1 if $BP_F$ is missing in HUNT3, 0 if present). Of $60385$ participants in HUNT2 without missing data on explanatory variables, $43.1 \%$ of the cohort were missing in HUNT3.

Summary statistics and units for the observations in the HUNT2 cohort are given in \cref{tb:summary_HUNT2}, together with observations grouped on missing status. 
In all analyses, $BP_F$, $BP_I$, $age$ and $BMI$ observations are standardized by the corresponding sample mean and standard deviation in the HUNT2 cohort.

\begin{table}
\centering
\caption{The sample mean and standard deviation
of $BP_F$, $BP_I$, $age$, and $BMI$ and proportion of female/male participants in the HUNT2 cohort are displayed in the third column. The fourth and fifth columns display sample mean for the present and missing participants in addition to proportions of present/missing participants for the whole cohort and per sex.} 
\begin{tabular}{llccc}
\multicolumn{5}{c}{\textbf{Summary of the HUNT2 cohort}} \\ \hline
Variable                                &Unit                       &HUNT2                              &Present in HUNT3                   &Missing in HUNT3    \\ \hline
\multicolumn{1}{l|}{$BP_3$ ($BP_F)$}    &\multicolumn{1}{l|}{mmHg}  & \multicolumn{1}{c}{-}             & 136.1             &   -                  \\
\multicolumn{1}{l|}{$BP_2$ ($BP_I$)}    &\multicolumn{1}{l|}{mmHg}  & \multicolumn{1}{c}{139.5 (23.6)}  & 135.2             & 145.0 \\
\multicolumn{1}{l|}{$age_2$}            &\multicolumn{1}{l|}{years} & \multicolumn{1}{c}{ 50.0 (17.1)}  & 47.0              & 54.01 \\
\multicolumn{1}{l|}{$BMI_2$}            &\multicolumn{1}{l|}{kg/$m^2$}& \multicolumn{1}{c}{~26.4 (~4.1)}& 26.2              & 26.6  \\
\multicolumn{1}{l|}{$sex$}              &\multicolumn{1}{l|}{}      &\multicolumn{1}{c}{}               & 56.9 \%           & 43.1 \%\\
\multicolumn{1}{l|}{\quad female}       &\multicolumn{1}{l|}{0} & \multicolumn{1}{c}{53.0 \%}           & 59.3 \%           & 40.7 \%\\
\multicolumn{1}{l|}{\quad male}         &\multicolumn{1}{l|}{1} & \multicolumn{1}{c}{47.0 \%}           & 54.3 \%           & 45.7 \%\\ \hline
\end{tabular}
\label{tb:summary_HUNT2}
\end{table}
\begin{figure}
    \centering
    \includegraphics{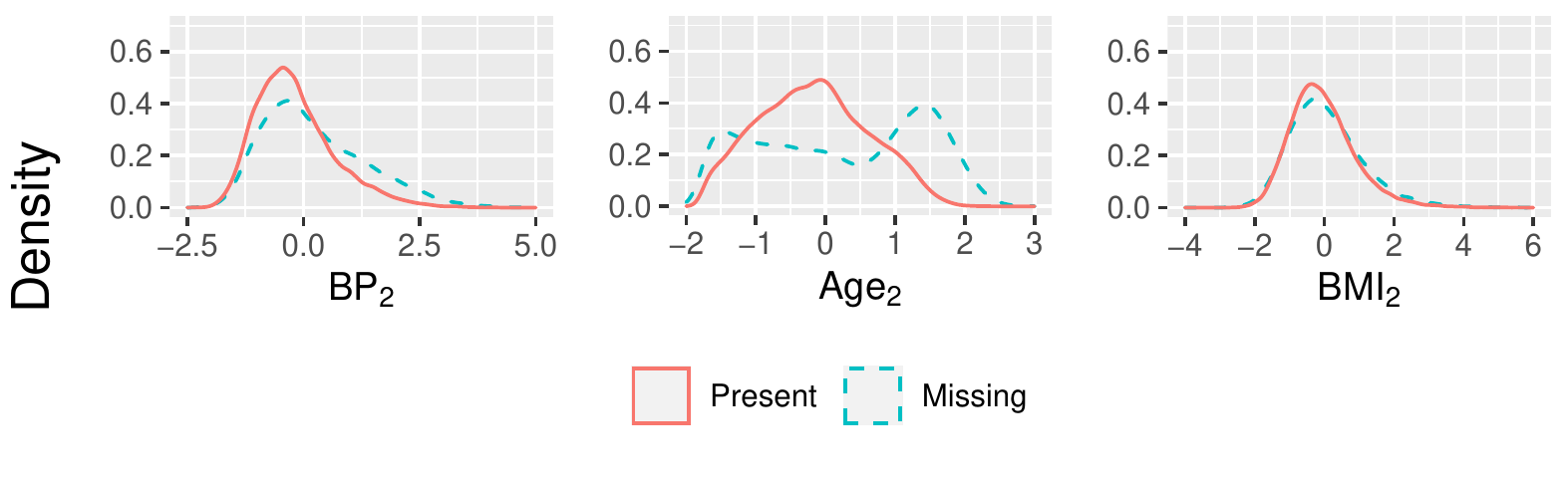}
    \caption{Smoothed empirical density of $BP_I$ ($BP_2$), $age$, and $BMI$ for all participants in the HUNT2 cohort.
    }
    \label{fig:smood_missing}
\end{figure}

In \cref{fig:smood_missing} and \cref{tb:summary_HUNT2} we find clear differences between the present and missing participants for $age$ and $BP_I$ ($BP_2$). Middle-aged participants are less likely to drop out than young or elderly participants, and those with higher blood pressure are more likely to be missing. This suggests that the data is at least MAR. 

We defined a validation dataset (HUNT3 cohort) consisting of participants with observations of blood pressure ($BP_I = BP_3$), $age$, $BMI$ and $sex$ in HUNT3, and future blood pressure ($BP_F=BP_4$) and missing status in HUNT4. Of $50201$ participants in HUNT3, $33.3 \%$ of the cohort dropped out before HUNT4. See \cref{tb:summary_HUNT3} in \cref{sec:summary_hunt3} for a summary of the HUNT3 cohort.

\subsection{A Shared Parameter Model for Blood Pressure}
\label{sec:spm}
We set up a shared parameter model (SPM) for $BP_F$ and the missing process using $age$, $sex$, $BMI$, and $BP_I$, as explanatory variables in the framework of a LGM as presented in \cref{sec:bg_latent_gaussion_models}. Let ${BP_F}_{i}$ and $m_i$ represent future blood pressure and missing status for individual $i$. The likelihoods are chosen to be Gaussian with identity link for $BP_F$ and Bernoulli with logit link for $m_i$; 
$BP_{F_i} \sim N(\eta_{BP_i}, \sigma_{BP}^2)$ and $m_i \sim Bernoulli(p_i)$ with $ logit(p_i) = \eta_{m_i}$. 
From the general formula of LGMs in \cref{sec:bg_latent_gaussion_models} the explanatory variables can be included either as linear effects or as non-linear effects. Based on the work by \cite{master_Lars} and the analyses 
in \cref{sec:spm_additive} we chose to include $age$ in the missing process as a non-linear effects, and all other explanatory variables as well as age in the blood pressure model as linear effects.
Further we introduce a shared parameter $\epsilon_i$ in both linear predictors, and an association parameter $c$ for the missingness model;
\begin{align}
\label{eq:spm}
    \eta_{BPi} = \alpha_0 + \alpha_{BP} {BP_I}_i + \alpha_{age} age_i + \alpha_{BMI} {BMI}_i + \alpha_{sex} sex_i + \epsilon_i\\ \nonumber
    \eta_{mi} = \beta_0 + \beta_{BP} {BP_I}_i + f(age_i) + \beta_{BMI} {BMI}_i + \beta_{sex} sex_i + c {\epsilon}_i,  \nonumber
\end{align}
where the shared parameters $\epsilon_i$ are assumed to be independent Gaussian $\epsilon_i \sim N(0,\sigma_{\epsilon}^2)$ and $f()$ is a random walk of order two with variance $\sigma_{age}$, as defined in \cref{sec:spm_additive}. 
To avoid identifiability issues between the shared parameter $\epsilon$ and the likelihood of $BF_F$ we fix $\sigma_{BP}^2$ to a small value ($\sigma_{BP}^2=0.001^2$). All regression parameters $\alpha_0$, $\alpha_{BP}$, $\alpha_{age}$, $\alpha_{BMI}$, $\alpha_{sex}$, $\beta_0$, $\beta_{BP}$, $\beta_{BMI}$, and  $\beta_{sex}$ are given independent priors $ N(0, {10^3}^2)$, and $\sigma_{age}^2$ and $\sigma_{\epsilon}^2$ are assigned independent gamma priors, $\text{Gamma}(1, 5 \cdot 10^5)$.  We expect the shared parameter to influence the missing process similarly or less than the standardized explanatory variables. Therefore, the association parameter $c$ is given an informative prior $c \sim N(0, 1^2)$. A sensitivity study is conducted for this prior, see \cref{sec:prior_sensitivity}. 

When the association parameter $c=0$, the models for $BP_F$  and $m$ are independent and we have a model that assumes data MAR. We refer to this model as the naive model, and it is used as a benchmark model. 

For simplicity we introduce some notation.
Let $\boldsymbol{x_i} = ({BP_I}_i, age_i, BMI_i, sex_i)$ be the explanatory variables and $\boldsymbol{y_i} = ({BP_F}_i, m_i )$ the response variables for individual $i$. Further let $ X = (x_i, .., x_n)^T$ be the explanatory variables for all $n$ participants and $Y = (y_1, ..., y_n)^T$ be the corresponding response variables in a cohort. When needed, we use a superscript to indicate the HUNT2 or HUNT3 cohort, i.e., $X^2$ are the explanatory variables from the HUNT2 cohort.
Denote the modeling parameters by $\boldsymbol{\theta}$ = ($\alpha_0$,  $\alpha_{BP}$,  $\alpha_{age}$, $\alpha_{BMI}$, $\alpha_{sex}$, $\beta_0$, $\beta_{BP}$, $\beta_{BMI}$, $\beta_{sex}$, $\epsilon$, $f$, $c$, $\sigma_{\epsilon}$, $\sigma_{age}$) where $f$ refer to the Gaussian variables of the additive effect for $age$.

\subsection{Inference}
\label{sec:inference}
Conditioned on data $(X,Y)$ we can achieve posterior distributions for the parameters, $\pi(\boldsymbol{\theta} | X,Y)$. In this work, we are either interested in the marginal posterior of selected parameters $(j)$, $\pi(\theta^{(j)} | X,Y)$, or in the posterior predictive distribution for a new person with explanatory variables $\boldsymbol{x_{new}}$. This posterior predictive distribution is given by 
 \begin{align*}
     \pi(\boldsymbol{y_{new}}|\boldsymbol{x_{new}},X, Y) &= \int \pi(\boldsymbol{y_{new}}, \boldsymbol{\theta}| \boldsymbol{x_{new}}, X, Y) d\boldsymbol{\theta} \\
     &= \int  \pi(\boldsymbol{y_{new}}| \boldsymbol{x_{new}}, \boldsymbol{\theta}) \pi(\boldsymbol{\theta}| X, Y) d\boldsymbol{\theta}. 
 \end{align*}

The SPM suggested in \cref{sec:spm} is a LGM, as described in \cref{sec:bg_latent_gaussion_models} that meets the requirements for using the computationally efficient integrated nested Laplace approximations (INLA), see \cite{steinsland2014quantitative} for more details for an analogous SPM.
The latent Gaussian field consists of ($\alpha_0$,  $\alpha_{BP}$,  $\alpha_{age}$, $\alpha_{BMI}$, $\alpha_{sex}$, $\beta_0$, $\beta_{BP}$, $\beta_{BMI}$, $\beta_{sex}$, $\epsilon$, $f$, $c$) and the non-Gaussian hyperparameters are ($\sigma_{\epsilon}$, $\sigma_{age}$).

\subsection{Validation Scheme Using the HUNT3 Cohort}
\label{sec:validation_prediction}

We evaluate the prediction models obtained from the HUNT2 cohort using the HUNT3 cohort.  For each participant $i$ in the HUNT3 cohort we get the predictive distributions $\boldsymbol{\hat{y}_{i}} \sim \pi(\boldsymbol{y_{i}^{3}}|\boldsymbol{x_{i}^{3}},X^{2}, Y^{2})$  and specifically for the future blood pressure and missing status $\hat{BP_{Fi}} \sim \pi(BP_{Fi}^{3}|\boldsymbol{x_{i}^{3}},X^{2}, Y^{2})$ and $\hat{m_{i}} \sim \pi(m_{i}^{3}|\boldsymbol{x_{i}^{3}},X^{2}, Y^{2})$. 

To evaluate the predictive performance we calculate the mean continuous rank probability score (CRPS) of $\hat{BP_{Fi}}$ and mean Brier score of $\hat{m_{i}}$ over all participants in the HUNT3 cohort. 
Let $F_{\hat{BP_{Fi}}}(x)$ be the cumulative probability distribution of $\hat{BP_{Fi}}$ and $BP_{Fi}$ the observed blood pressure in HUNT4 for participant $i$, then $CRPS(F, y) = \int_{-\infty}^{\infty}[F(x) - H(x-BP_{Fi})]^2 dx$ where $H(u)$ is the Heaviside function (0 for $u<0$ and 1 for $u>0$).
The blood pressure model can only be validated on the participants observed in both surveys. In contrast, the missing model can be evaluated for all participants.

Predictions from the SPM and the naive model are compared by their posterior mean for the HUNT3 cohort participants as well as their CRPS and Brier scores.


\subsection{Evaluation of Missing not at Random by Conditioning on Missing Status}
\label{sec:validation_mnar}

We introduce a novel method for validating if data is MNAR based on a SPM fitted to a training dataset (the HUNT2 cohort) and the difference in predictive performance for a validation dataset (the HUNT3 cohort) for predictors with and without conditioning on the missing status in the training dataset. For readability, we introduce the method using the notation of HUNT2 cohort and HUNT3 cohort, but the method is general. 

If data are MNAR and the SPM is true, there is information about the shared parameter in the missing status, and conditioning on the missing status, i.e. the value of $m_{new}$ should give a better predictor.  For each participant $i$ in the HUNT3 cohort we can, from the predictive distribution, $\pi(\boldsymbol{y_{i}^{3}}|\boldsymbol{x_{i}^{3}},X^{2}, Y^{2})$  derive both the marginal predictive distribution for the future blood pressure  $\hat{BP_{Fi}} \sim \pi(BP_{Fi}^{3}|\boldsymbol{x_{i}^{3}},X^{2}, Y^{2})$ and the  predictive distribution for the future blood pressure conditioned on the missing status $\hat{BP_{Fi}} | m_{i} \sim  \pi(BP_{Fi}|\boldsymbol{x_{i}^{3}},X^{2}, Y^{2}, m_i^3)$. 

In practice, we can for a validation dataset only evaluate the predictions for the present participants, and not the dropouts, and we therefore compare the predictive performance of $\hat{BP}_{F_{new}} $ and $\hat{BP}_{F_{new}} | m_{new}=0$ for all presents participants. In this work, we have calculated the absolute error of the posterior mean predictions for each participant, and compare mean absolute errors (MAE) for the prediction with and without conditioning on missing status.

\subsection{Software and Code}

In this work, we use the R-INLA software \citep{r-inla}. 
The R-INLA software supports fitting models with multiple likelihoods \cite[Chap. 6.4]{steinsland2014quantitative, master_Lars, gomez2020bayesian} which is the case for the SPM \cref{eq:spm}.
All the code 
is available at the GitHub repository \cite{git_repo}. Since data can not be shared, to protect participants privacy, a completely simulated dataset is provided at \cite{git_repo}.

\section{Results for the Blood Pressure Case Study}
\label{sec:results}

The shared parameter models (SPM)
and the naive model 
introduced in \cref{sec:m&m} are fitted using the HUNT2 cohort as described in \cref{sec:spm}. This chapter presents and compares the posterior distributions of interest. 
Further, the predictive models are evaluated through predictive performance of the HUNT3 cohort, as described in \cref{sec:validation_prediction} and \cref{sec:validation_mnar}. 

\subsection{
Results for 
the HUNT2 Cohort}
\label{sec:results_param_est}

\begin{figure}
    \centering
    \includegraphics[width = \textwidth]{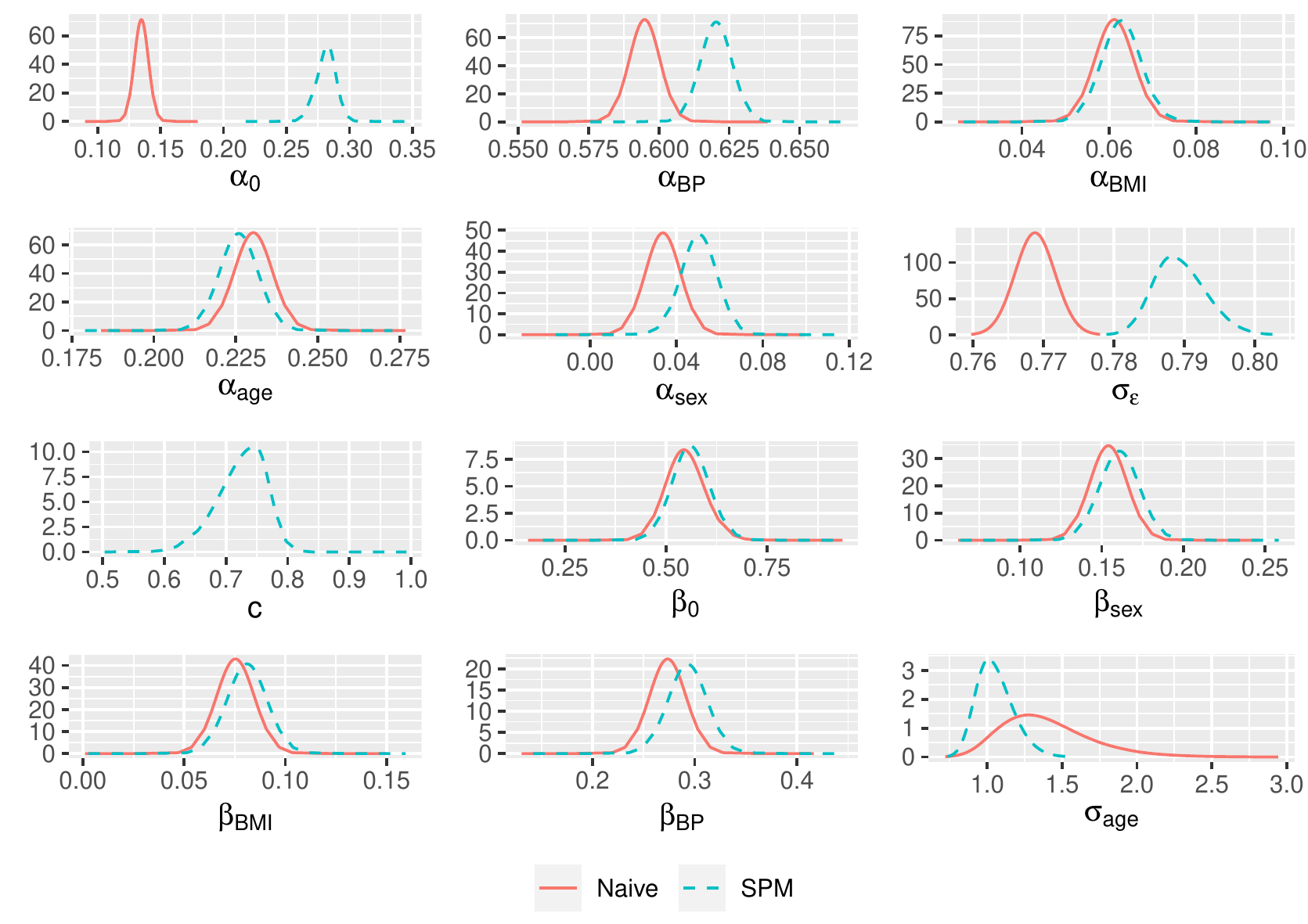}
    \caption{Posterior distribution of the latent field and hyperparameters for the SPM 
    and naive model 
    fitted to the HUNT2 cohort.}
    \label{fig:post_scaled}
\end{figure}

The posterior distributions of the estimates obtained by the SPM 
and the naive model
, introduced in \cref{sec:spm} and fitted to the HUNT2 cohort, can be seen in \cref{fig:post_scaled}. The posterior mean and $95 \%$ credible intervals are presented in \cref{tb:summary_param_est} in \cref{apendix:param_est}. 

For the blood pressure submodel in the SPM, we see that
the effect of $BP_I$ ($\alpha_{BP}$) is the largest followed by $age$, $BMI$ and $sex$. 
The effect of $BMI$ and $sex$ are close to zero. The parameter estimates of the naive 
blood pressure model are of the same order as the SPM. However, all variables but age have weaker effects on future blood pressure in the naive model than the SPM. The difference is especially pronounced for $\alpha_{0}$ and $\alpha_{BP}$, suggesting the two models could result in different predictions. 

\begin{figure}
    \centering
    \includegraphics{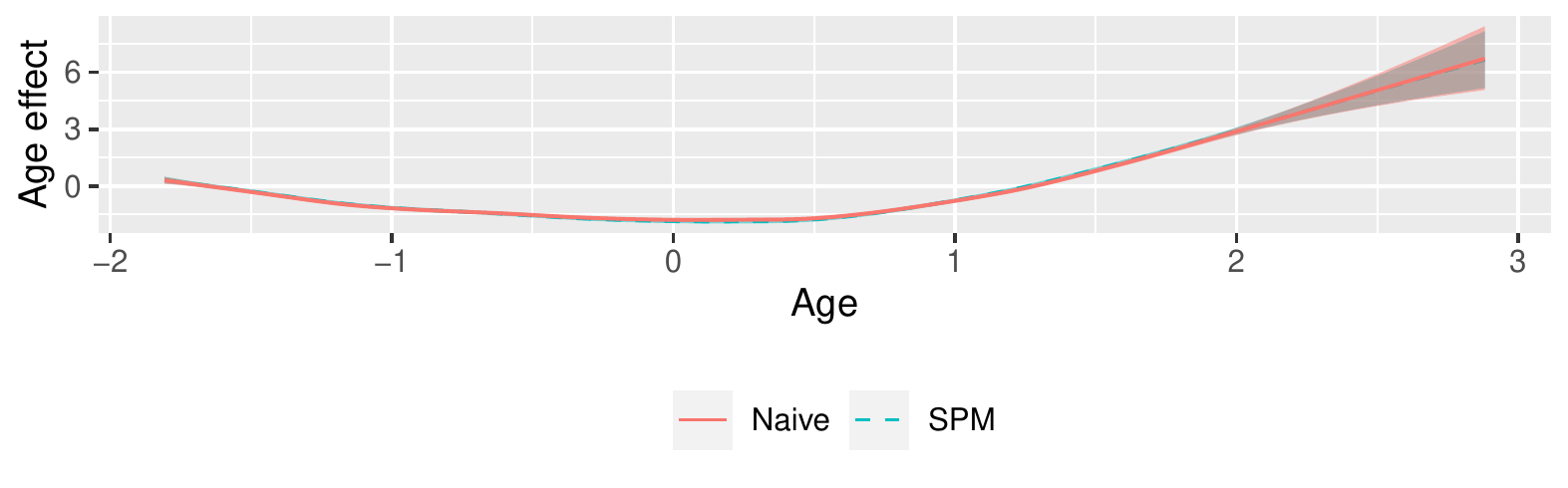}
    \caption{Age effect for the SPM and naive model fitted to the HUNT2 cohort with $95\%$ credible bands.}
    \label{fig:age_effect}
\end{figure}

For the missing process in the SPM, we see that $sex$ has the largest effect on the probability of dropping out, followed by $BP_I$ and $BMI$. The age effect is largest for the elderly and smallest for middle-aged participants (\cref{fig:age_effect}). 
The SPM and the naive model are more similar in parameter estimates for the missing process than the blood pressure process. However, the parameter estimates of the naive model are shifted towards lower values than the SPM.

The association parameter $c$, connecting the two submodels \cref{eq:spm} in the SPM is clearly positive, which implies an increase in the probability of dropping out for larger random effect. Hence, according to the SPM, participants with higher $BP_F$ than expected from the explanatory variables also have a larger probability of dropping out than expected from the explanatory variables.

\subsection{Results for Toy Example Participants}

\begin{figure}
    \centering
    \includegraphics{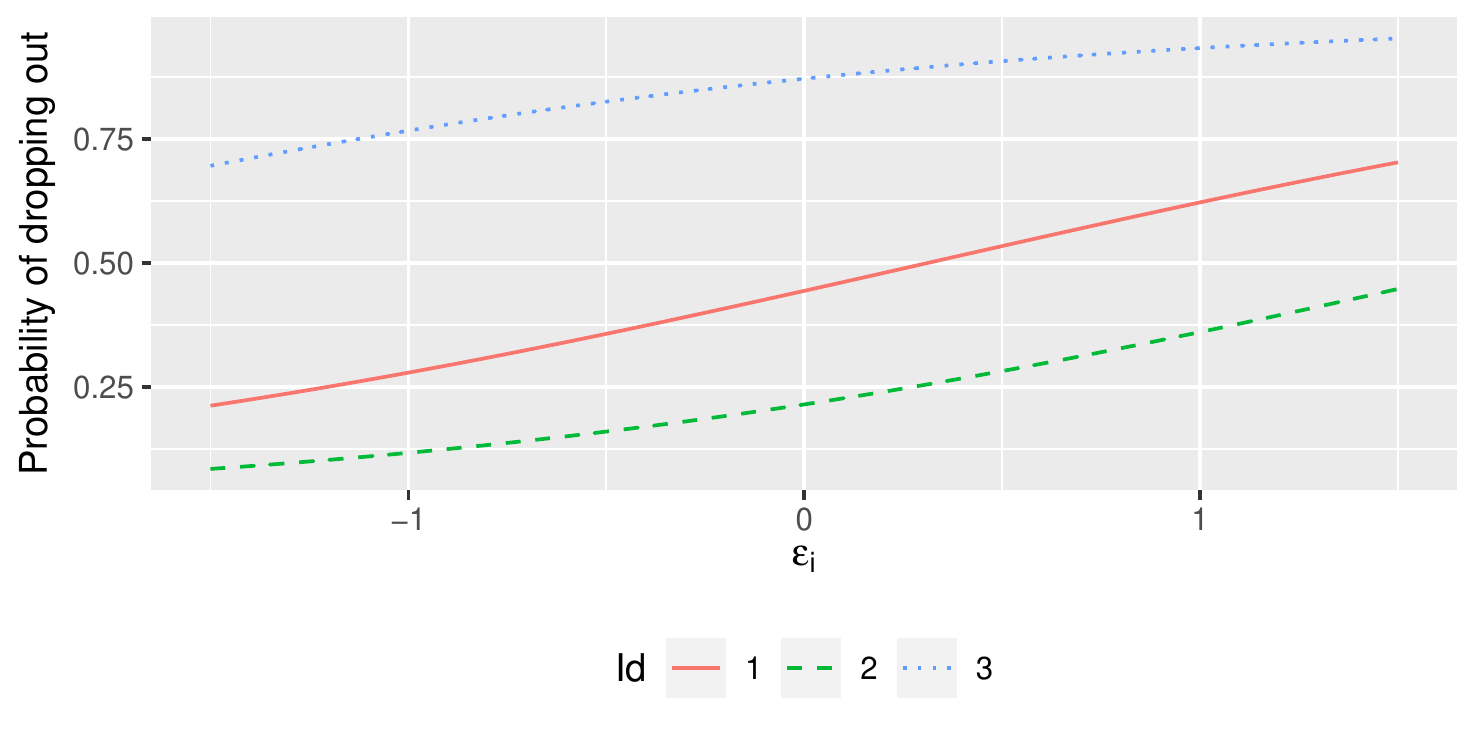}
    \caption{The probability of dropping out as a function of the individual random effect of the blood pressure model for three toy example participants specified in \cref{tb:individuals}.}
    \label{fig:assosiation}
\end{figure}

Since the data used in this work contains personal information, we consider three constructive toy example participants to explore the model presented in \cref{sec:spm} and \cref{sec:validation_prediction} on an individual level.  We used a young and underweight female with low $BP_I$ (id1), a middle-aged and overweight female with average $BP_I$ (id2), and an old and obese female with severely high $BP_I$ (id3), see \cref{tb:individuals}. For these participants, the effect of the association parameter on the probability of dropping out is plotted in \cref{fig:assosiation} for random effects between $-1.5$ and $1.5$ which corresponds  to approximately two standard deviations of the random effect. We find that a larger value of the random effect gives a large probability of dropping out for all three toy example participants.

\begin{table}
\center
\caption{Values of $BP_I$, $age$ and $BMI$ for three  female toy example participants.}
\begin{tabular}{l|lllllll}
id & $BP_I^*$  & $age^*$  & $BMI^*$  & $sex$ & ${BP_I}_{true}$ & $age_{true}$ & $BMI_{true}$\\ \hline
1  & -2 & -1.5    & -2 & \text{female}  & 92.2  & 24.4 & 18.2 \\
2  & 0    & 0   & 0    &  \text{female} & 139.5 & 50.0 & 26.4\\
3  & 2  & 1.5     & 2  & \text{female}  & 186.7 & 75.7 & 34.6 \\
\multicolumn{8}{l}{\small{* Standardized values}}
\end{tabular}
\label{tb:individuals}
\end{table}

\begin{figure}
    \centering
    \includegraphics{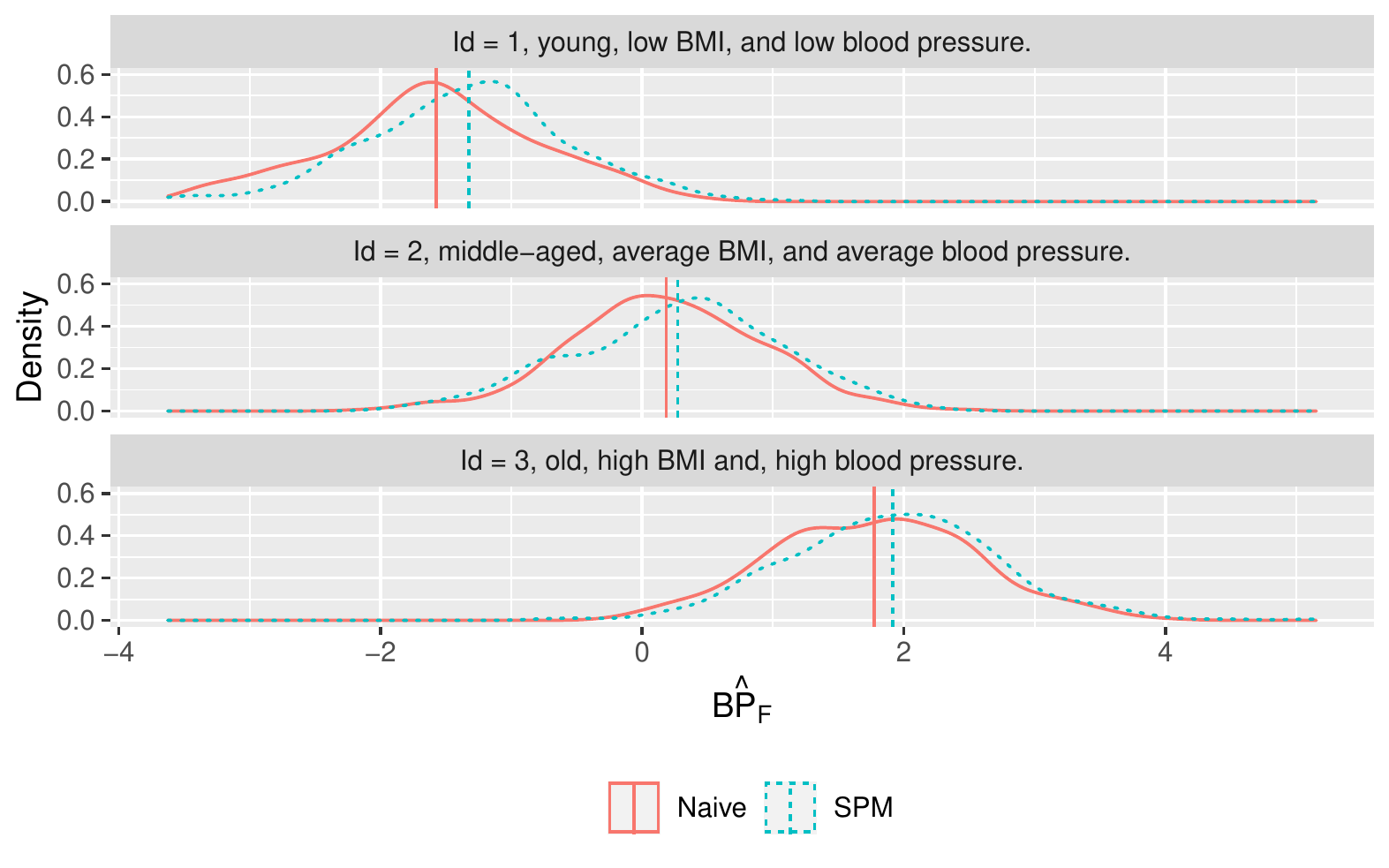}
    \caption{Posterior predictive distribution of the future blood pressure ($BP_F$) for three toy example participants in \cref{tb:individuals} for the naive model and the SPM models. Vertical lines indicate posterior means. 
    }
    \label{fig:individual_bp}
\end{figure}

The posterior predictive distributions of $BP_F$ for the three simulated participants are plotted in \cref{fig:individual_bp} for both the SPM and the naive model. 
For all toy example participants, the posterior predictive distribution from the SPM is shifted towards larger values than the naive model, and more so for id 1 and 3 who have more extreme explanatory variables. 

\subsection{Validation of Model Predictions for the HUNT3 Cohort}

\begin{table}[h]
\center
\caption{CRPS score for predictions of systolic blood pressure in HUNT4 for the present participants. Brier score for mean predictions of the probability of drop out in HUNT4. All predictions are based on the HUNT3 cohort. The best score is indicated in bold.}
\begin{tabular}{l|c|ccc}
              & \textbf{CRPS}       &\multicolumn{3}{c}{\textbf{Brier}} \\
\textbf{}      &                    &All       & Present    & Missing   \\ \hline
\textbf{SPM}   & 0.4406             &0.2082    & 0.1656     & \textbf{0.2937}    \\
\textbf{Naive} & \textbf{0.4337}    &\textbf{0.2072}    & \textbf{0.1602}     & 0.3014   
\end{tabular}

\label{tb:scores}
\end{table}

\begin{figure}
    \centering
    \begin{subfigure}[b]{1\textwidth}
    \centering
    \includegraphics{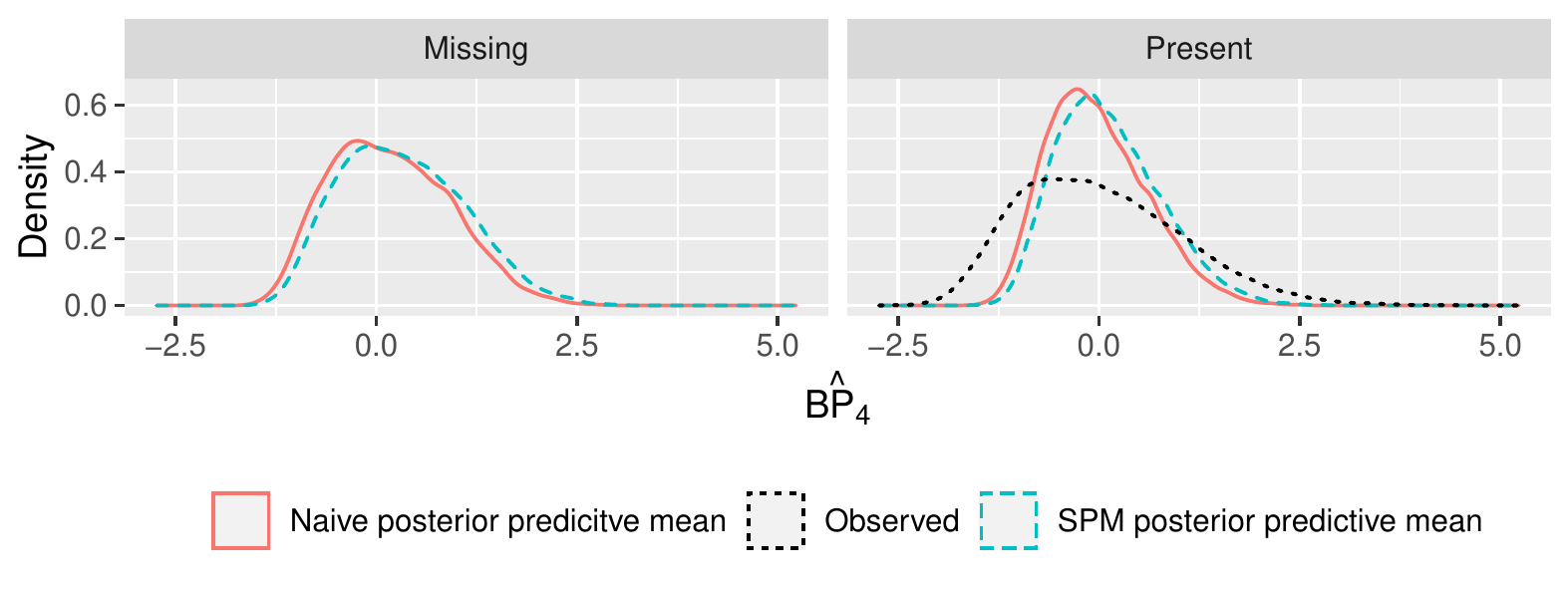}
    \caption{Density over all participants in HUNT3 of the posterior mean of predicted future blood pressure $\hat{BP_F}_i$.} 
    \label{fig:pred_bp_3_smooth}
    \end{subfigure}
    \begin{subfigure}[b]{1\textwidth}
    \centering
    \includegraphics{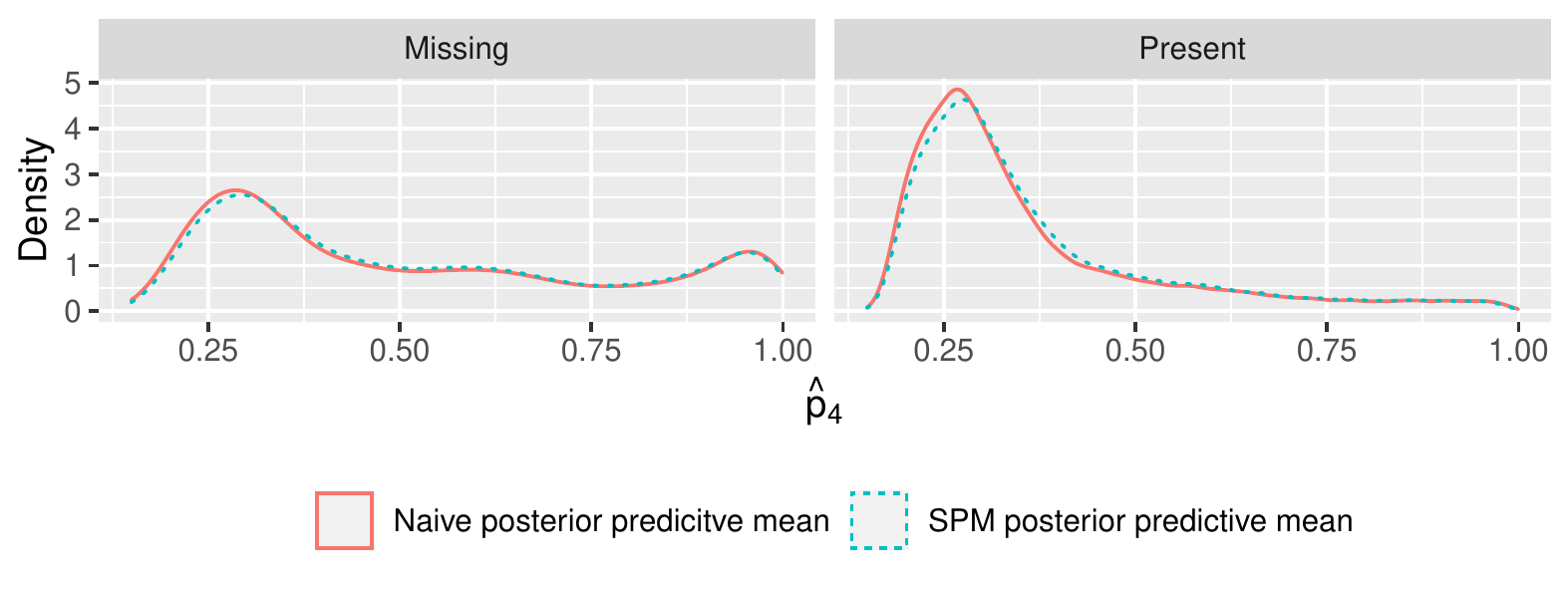}
    \caption{Density over all participants in HUNT3 of the posterior mean of predicted probability of dropping out $\hat{p}_i$.}
    \label{fig:pred_missing_smooth}
    \end{subfigure}
    \caption{All plots display the participants grouped by missing status in the HUNT3 cohort.}
\end{figure}

For all HUNT3 participants, the predictive distributions are calculated as described in \cref{sec:inference} using a model trained on the HUNT2 cohort.
The empirical distributions of posterior mean predictions for participants in HUNT3 cohort are found in \cref{fig:pred_bp_3_smooth}, and for the $BP_F$ and in \cref{fig:pred_missing_smooth} for probability of drop out $p$.
For reference the empirical distribution for observed  $BP_F$ is included for present participants in \cref{fig:pred_bp_3_smooth}. 
We see that on the SPM predicts slightly larger values for the $BP_F$ than the naive model on the population level. 
The distributions of mean predictions of the probability of dropping out are very similar. The SPM predicts a slightly higher average probability of dropping out than the naive model. We note that the differences between naive and SPM have little practical implication on population level, but for some individuals there can be differences of some practical significance. 

We further evaluate the predictive performance for the HUNT3 cohort for the SPM and the naive model from the HUNT2 cohort as described in \cref{sec:validation_prediction}.
Mean CRPS for $BP_F$ and mean Brier scores for the probability of drop out are given in \cref{tb:scores}. The mean CRPS is very similar for the SPM and naive model, but the CRPS for the naive model predictions is slightly smaller and hence the naive model performs slightly better. This might be explained by the fact that the missing participants do not affect the likelihood of the naive model. Hence, the model is optimized to perform well for the present participants. We explore this further in a simulation study  in \cref{sec:simulation_validation}.

The Brier scores in \cref{tb:scores} are slightly better for the naive model than for the SPM when evaluating for all participants. However, when grouped by missing status, the naive model performs better on the present participants, and the SPM performs better on the dropouts.

\subsection{Evaluate of the Missing Not at Random Assumption}

We evaluate the MNAR assumption by comparing the predictive performance of the SPM with and without conditioning on the missing status for the HUNT3 cohort as described in \cref{sec:validation_mnar}.
The results are given in \cref{tb:validatoin_mnar}. We see that the mean absolute errors (MAEs) are very similar, but that the predictions for $\boldsymbol{BP_4}$ given $\boldsymbol{m_4}$ yield a slightly smaller error than not knowing  $\boldsymbol{m_4}$. This suggests that some information from the missing process affects the $\boldsymbol{BP_F}$. A simulation study is conducted in \cref{sec:simulation_validation}, and a difference of $-0.0014$ is within what is to be expected when the SPM is true for similar training and validation datasets.

\begin{table}
\center
\caption{Mean absolute error for $\boldsymbol{\hat{BP}_F} | m$ and $\boldsymbol{\hat{BP_F}}$. The best score is indicated in bold.}
\begin{tabular}{l|c|c|c}
\textbf{} & $\boldsymbol{\hat{BP}_F} | m$ & $\boldsymbol{\hat{BP_F}}$ & MAE($\boldsymbol{\hat{BP}_F} | m$) - MAE($\boldsymbol{\hat{BP_F}}$) \\ \hline
  \textbf{MAE}       &   \textbf{0.6140}         &   0.6155        &     -0.0014     
\end{tabular}
\label{tb:validatoin_mnar}
\end{table}
\section{Simulation Studies}
\label{sec:simulation_study}

We set up several simulation studies to explore the properties of the SPM, the naive model and the method validating MNAR by conditioning on missing status for datasets with the size and structure of the HUNT2 and HUNT3 cohort. In all simulation studies data sets for training models are simulated using the same number of participants and explanatory variables as in the HUNT2 cohort, i.e. $X=X^{2}$. Further, when simulating data, the parameters $\boldsymbol{\theta}= \boldsymbol{\theta^{true}}$ are set to the posterior mean estimates of the SPM and naive model fitted on the HUNT2 cohort, see \cref{tb:summary_param_est}, in \cref{apendix:param_est}, in all simulation studies. When predictions are studied, simulated validation datasets are based on the same size and explanatory variables as the HUNT3 cohort, i.e. $X=X^{3}$ is used.

\subsection{Simulation Study exploring Bias and Coverage}
\label{sec:sim_identifiability}

The aim of this simulation study is to study the properties of the posterior estimates for the SPM and the naive model in a situation similar to the HUNT2 cohort. 100 independent new response data (i.e. both future blood pressure $\boldsymbol{BP_F}$ and the missing status $\boldsymbol{m}$) are simulated using the SPM with parameters $\boldsymbol{\theta^{true}}$ and with explanatory variables as in the HUNT2 cohort, resulting in $Y^{(l)}$ for $l=1 \dots 100$. For each of the data sets $(X^2,Y^{(l)})$ both the SPM and the naive model are fitted. This gives posterior distributions $\pi(\boldsymbol{\theta} | X^2, Y^{(l)})$. Each of these are summarized by the posterior mean and the coverage indicator of the true value in $95 \%$ credibility interval.  

The resulting mean posterior mean, bias, and coverage are found in \cref{tb:sim_mnar_spm_est} in \cref{sec:supplementary_simulation_identifiability}.
\begin{figure}
    \centering
        \includegraphics[width =\textwidth]{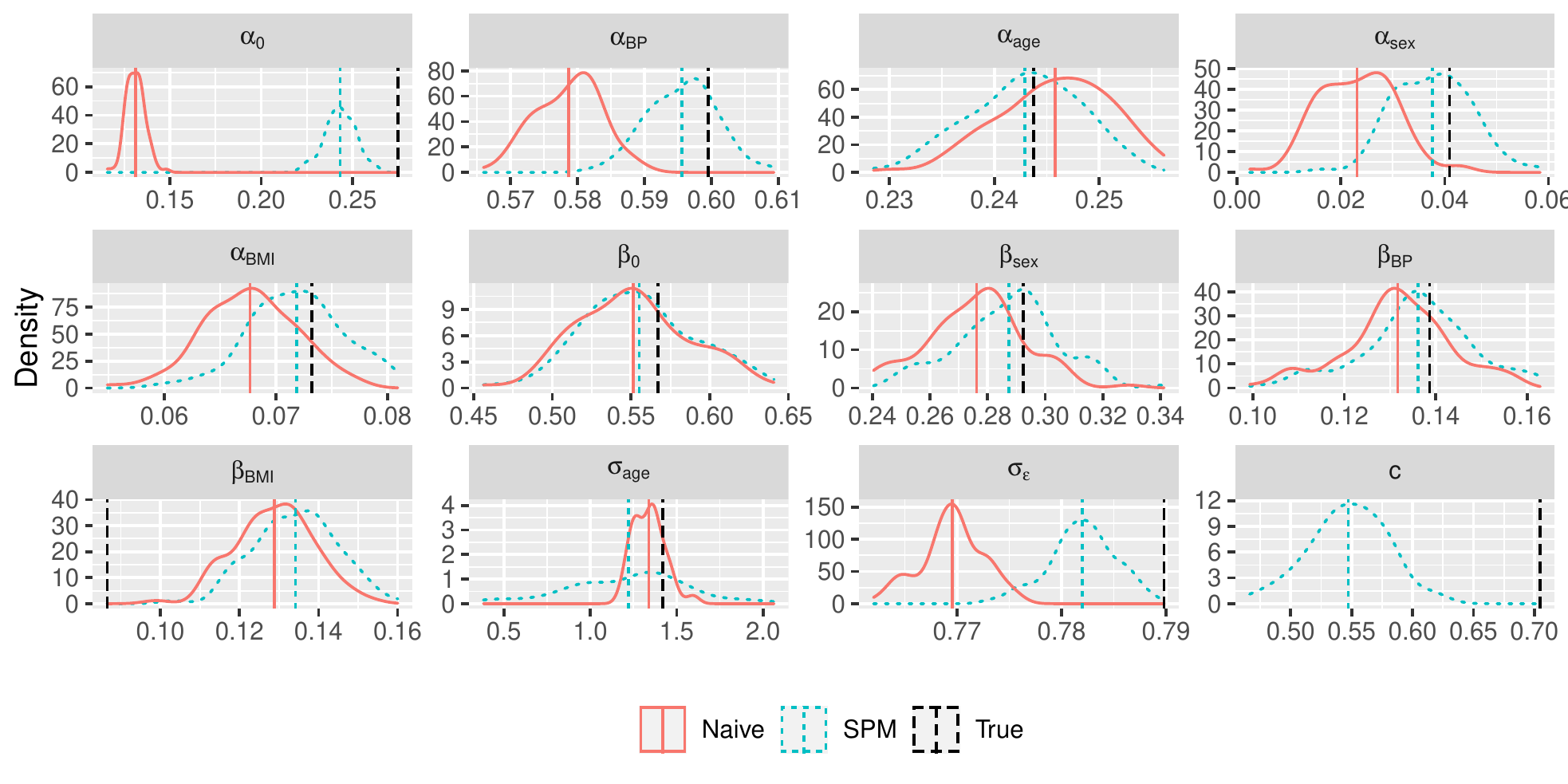}
        \caption{Distribution of posterior mean estimates on simulated data MNAR following the SPM as described in \cref{sec:sim_identifiability}. The mean of posterior means for both the SPM and naive model and the true value are indicated by the vertical lines. }
        \label{fig:simulation_study_MNAR_spm}
\end{figure}
The distribution of the posterior means from the simulation study when the true parameters are the posterior mean estimates of the SPM can be seen in \cref{fig:simulation_study_MNAR_spm}. 
From \cref{fig:simulation_study_MNAR_spm} we find  that both the naive model and SPM are biased when the data is generated from the SPM. However, the SPM is less biased especially for the blood pressure model parameters (i.e. $\alpha_0, \alpha_{PB}, \alpha_{age}, \alpha_{sex}$, and $\alpha_{BMI}$).  The association parameter c has especially low coverage, see \cref{tb:sim_mnar_spm_est} in \cref{sec:supplementary_simulation_identifiability}).

We have also preformed a similar simulation study, but with data MAR by setting the association parameter $c=0$ when simulating data. The results are presented in \cref{sec:supplementary_simulation_identifiability} When the data is MAR, both the SPM and naive model have very little bias, and in particular the association parameter $c$ is centered around zero and has good coverage.

\subsection{Simulation Study for Predictions}
\label{sec:simulation_validation}

The aim of this simulation study is to learn about the predictive performance. Following the procedure in \cref{sec:sim_identifiability} a data sets mimicking the HUNT2 cohort, $(X^{2},Y)$, and  $l=1 \dots 100$ data sets mimicking the HUNT3 cohort, $(X^{3},Y^{(l)})$, are simulated using the SPM with parameters $\boldsymbol{\theta^{true}}$. Corresponding posterior distributions based on the simulated HUNT2 cohort $\pi_{SPM}(\boldsymbol{\theta} | X^{2}, Y)$ and $\pi_{naive}(\theta | X^{2}, Y)$ are found based on the SPM and the naive model, respectively. From these, posterior predictive distributions are achieved, and the predictive performance is evaluated for each participant in the simulated data set $(X^{3},Y^{(j)})$ as described in \cref{sec:validation_prediction} by the CRPS and Brier score. A large advantage for the simulated data sets is that predictive performance can be evaluated not only for present participants, but also for those that are missing, and we include them.

Further we compare the predictive performance conditioned on missing status \\ 
$\pi_{SPM}(Y^{(l)} |\boldsymbol{m}^{l}, X^{3}, \boldsymbol{\theta})$ with $\pi_{SPM}(Y^{(l)} |X^{3}, \boldsymbol{\theta})$ through MAE as described in \cref{sec:validation_mnar}.

\begin{figure}
    \centering
    \begin{subfigure}[b]{0.45\textwidth}
    \centering
    \includegraphics[width = \textwidth]{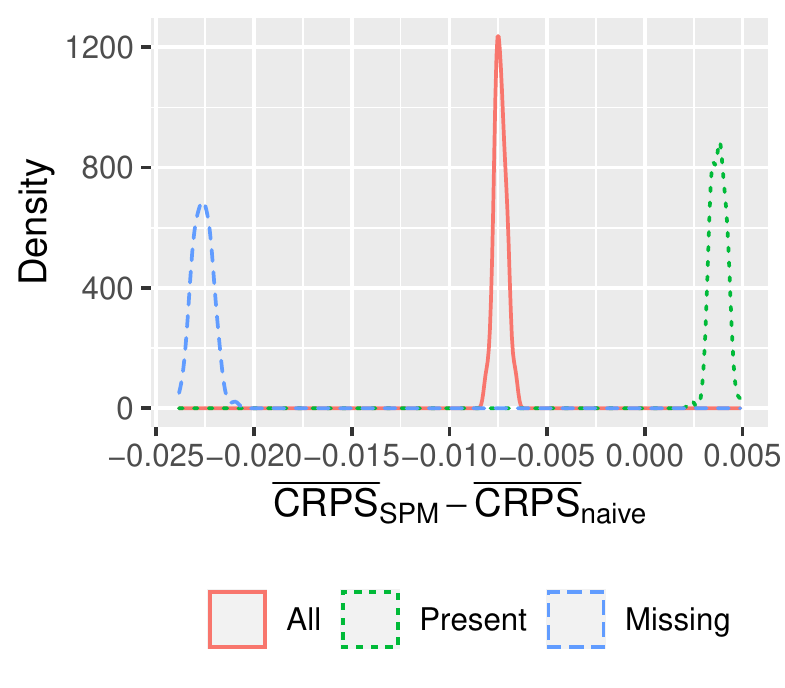}
    \caption{Difference in mean CRPS score for the SPM and naive model with 100 simulations} 
    \label{fig:diff_crps}
    \end{subfigure}
    ~
    \begin{subfigure}[b]{0.45\textwidth}
    \centering
    \includegraphics[width = \textwidth]{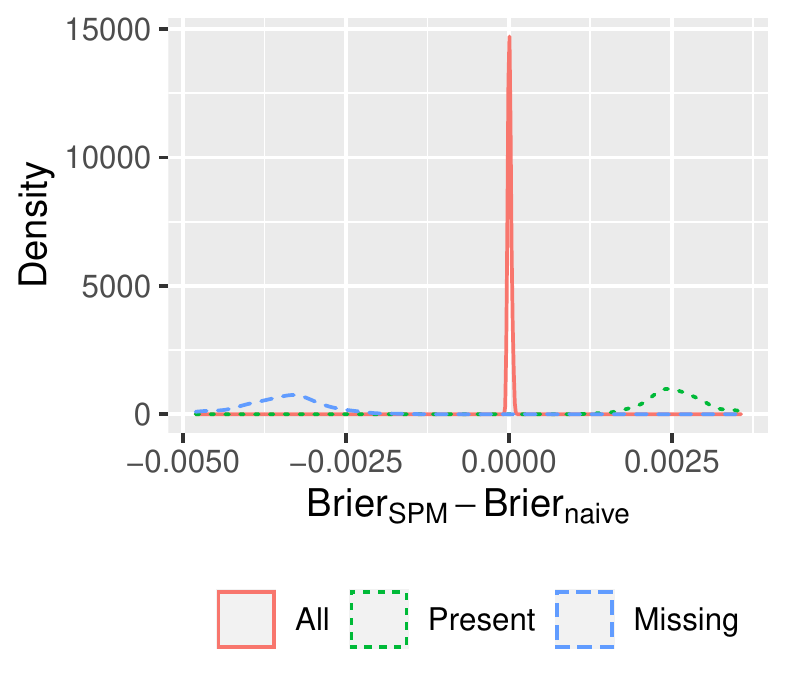}
    \caption{Difference in mean Brier score for the SPM and naive model with 100 simulations}
    \label{fig:diff_brier}
    \end{subfigure}
    \caption{Plots for differences in mean scores for participants grouped by missing status in the HUNT3 cohort.}
\end{figure}

\Cref{fig:diff_crps} displays the distribution of the difference between the mean CRPS for the SPM and the naive model. This difference is displayed for all simulated participants (present/missing) and grouped on missing status. We see that the SPM performs better for all participants and the dropouts, but the naive model performs better on present participants. 
This demonstrates that for our case study even if the data is MNAR and follows the SPM, the naive model is expected to obtain a better CRPS than the SPM when only  evaluating for present participants (and the missing participants can not be used for evaluation!).

\Cref{fig:diff_brier} shows the distribution of the difference in Brier score for all, missing and present  participants.
We see similar results here, 
the SPM predicts best for missing participants and the naive model predicts best for the present participants.

\begin{figure}
    \centering
    \includegraphics{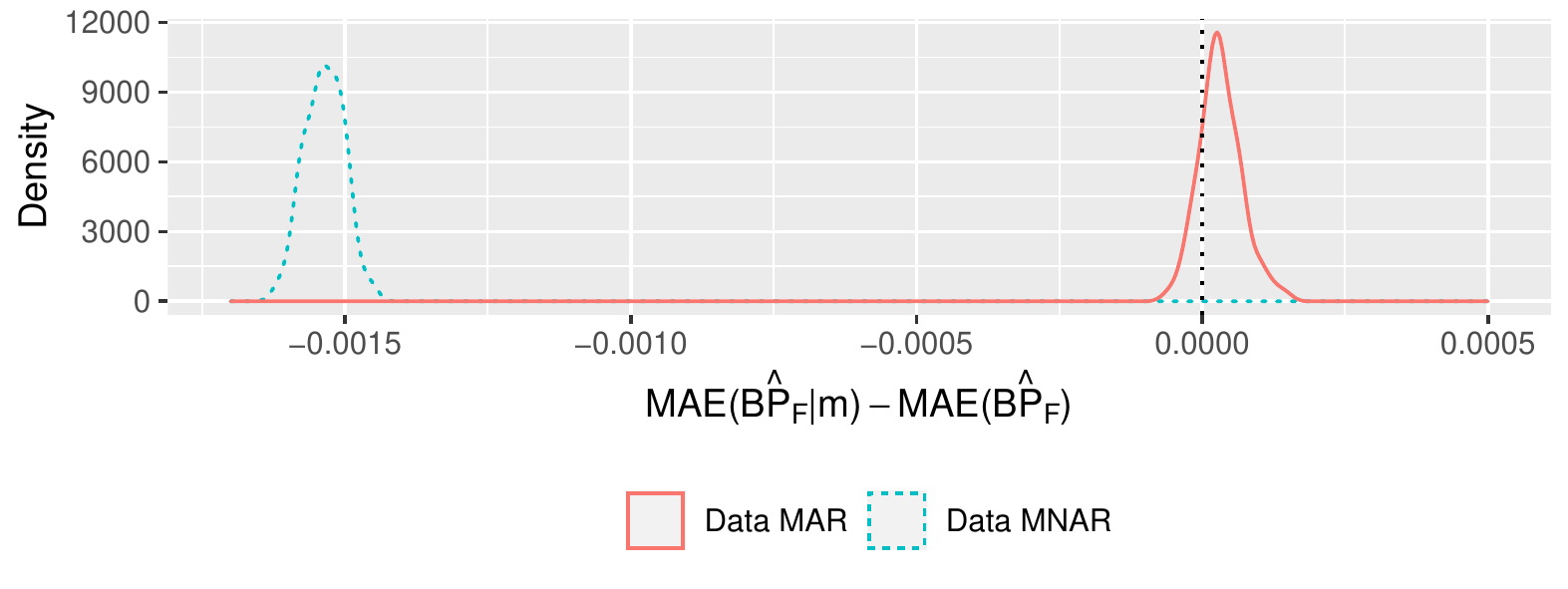}
    \caption{Difference in mean absolute error (MAE) of the predictions of $\boldsymbol{\hat{BP_F}|m}$ and $\boldsymbol{\hat{BP_F}}$ when the data is MNAR and MAR. The vertical line indicates zero. }
    \label{fig:validation_mnar_diff}
\end{figure}

\begin{figure}
    \centering
    \begin{subfigure}[t]{0.45\textwidth}
        \includegraphics[width = \textwidth]{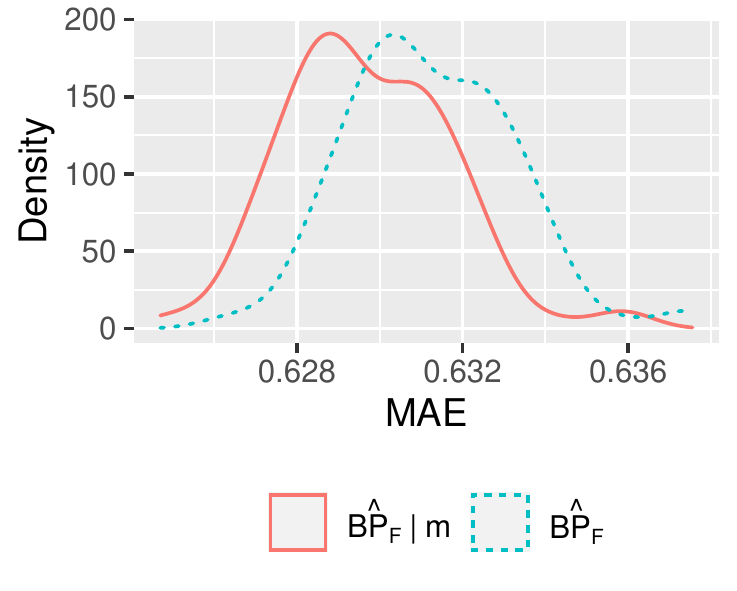}
    \caption{ 
    Data MNAR}
    \label{fig:validation_mnar_spm}
    \end{subfigure}%
    ~
    \begin{subfigure}[t]{0.45\textwidth}
        \includegraphics[width = \textwidth]{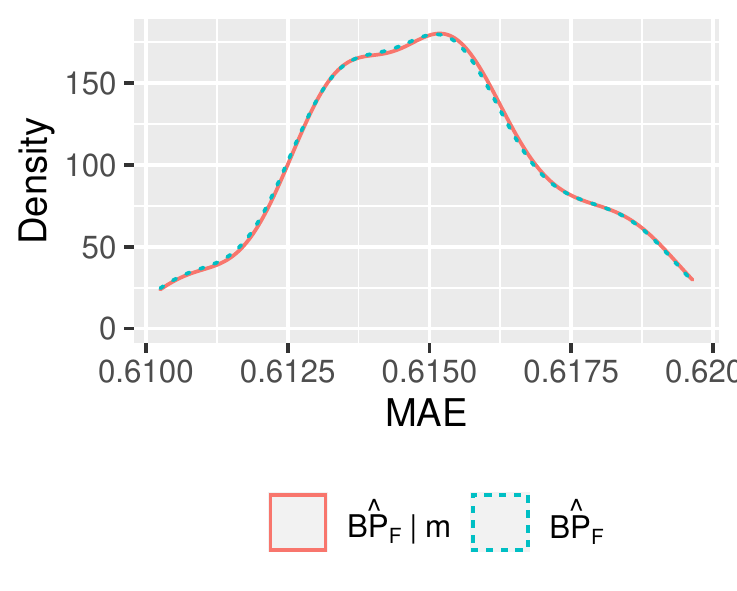}
    \caption{ 
    Data MAR}
    \label{fig:validation_mnar_naive}
    \end{subfigure}
    \caption{Distribution of the mean absolute error MAE of the posterior mean prediction for future blood pressure obtained from the validation scheme presented in \cref{sec:simulation_validation}}
\end{figure}

For data MNAR, we can see from \cref{fig:validation_mnar_spm} that the distribution of MAEs for the 100 simulations is shifted towards lower values when the missing status is known when the data is MNAR. Further, we see from \cref{fig:validation_mnar_diff} that the MAE was smaller for every simulated dataset when predicting $\boldsymbol{\hat{BP_F}}|\boldsymbol{m}$ than $\boldsymbol{\hat{BP_F}}$ since zero is not contained in the distribution of $MAE(\boldsymbol{\hat{BP_F|m}}) -MAE(\boldsymbol{\hat{BP_F}})$. 
Hence, for data MNAR following the SPM, we can expect the predictions of $\boldsymbol{\hat{BP_F|m}}$ to be better than $\boldsymbol{\hat{BP_F}}$. 

For data MAR \cref{fig:validation_mnar_naive} shows we obtain almost the same mean absolute error for every simulated dataset when predicting $\boldsymbol{\hat{BP_F}}|\boldsymbol{m}$ as when predicting $\boldsymbol{\hat{BP_F}}$. In addition, we see from \cref{fig:validation_mnar_diff} that the distribution of $MAE(\boldsymbol{\hat{BP_F}|m}) - MAE(\boldsymbol{\hat{BP_F}})$ covers zero. 

From \cref{fig:validation_mnar_diff}, we see that there is no overlap between the two distributions for, $MAE(\boldsymbol{\hat{BP_F|m}}) -MAE(\boldsymbol{\hat{BP_F}})$.  Also, the distributions are narrow both when the data is MNAR and MAR. This demonstrates that even minor differences in MAE between predictions of $\boldsymbol{\hat{BP_F}|m}$ and $\boldsymbol{\hat{BP_F}}$ indicate that data are MNAR.

\section{Discussion}
\label{sec:discussion}

In this work, we propose reasonable models in the form of a shared parameter model (SPM)
for predicting future blood pressure ($BP_F$) and missing status based on the HUNT Study accounting for missing data in the response.
The SPM is compared to a naive model which assumes data MAR. There are some, but not massive, differences between the two models. Especially the effect of current blood pressure is larger in the SPM than the naive model. 
The simulation study confirms that if the underlying assumption of data following the SPM is valid, the SPM accounts for this better than the naive model. 
We note that there seem to be some issues resulting in a bias when the data is MNAR and a low coverage for some model parameters in the simulation studies (\cref{fig:simulation_study_MNAR_spm}, \cref{fig:simulation_study_MAR_spm}, \cref{tb:sim_mnar_spm_est}, and \cref{tb:sim_mar_spm_est}). This bias is especially pronounced for the association effect $c$ when the data is MNAR. Initially we suspected that the low coverage could be due to misspecification of the informative prior used for $c$. However, the prior sensitivity study presented in \cref{sec:prior_sensitivity} indicates that the model is robust to prior specifications concerning the association parameter.

In this paper we have validated the models based on predictive performance based on a validation data set, both in straight forward predictive performance for the present participants and through the new validation scheme proposed in \cref{sec:validation_mnar}. The results indicate that missing status contains information about future blood pressure ($BP_F$) given the SPM. According to the simulation studies on this validation scheme (\cref{sec:simulation_validation}), the model predictions conditioned on missing status are only better than those without knowledge on missing status if data is MNAR (\cref{fig:validation_mnar_diff}). When the data is MAR, the mean absolute error of the two posterior mean predictions are almost identical. Hence, we have strong indications that blood pressure missing in the HUNT Study is MNAR.

As this research was motivated by the need of a predictive model of future blood pressure accounting for missing data, we performed all simulation studies related to bias, coverage and predictive performance on data sets that mimic our study system in size and explanatory variables, i.e. the HUNT2 cohort for training and the HUNT3 cohort for validation.
Exploring asymptotic properties of the models and validation schemes or how uncertainty, biases and predictive performance change with the size of the datasets and missingness have been outside the scope of this work.   

Driven by the results of this case study we find the properties of the validation schemes an interesting topic of further research. One approach would be to 
study the models and results of this work in the framework of missingness graphs introduced by \cite{mohan2021graphical} and the missing at random counterpart models of \cite{molenberghs2008every}.

We acknowledge that the models presented in this work are non-trivial to set up and are computationally demanding. With the available code it is relatively straightforward to do inference and simulation studies for similar datasets. The quantity of interest do not need to follow a Gaussian likelihood, but can be any member of the exponential family.

\section{Acknowledgement}

The Trøndelag Health (HUNT) Study is a collaboration between HUNT Research Centre (Faculty of Medicine and Health Sciences, Norwegian University of Science and Technology NTNU), Trøndelag County Council, Central Norway Regional Health Authority, and the Norwegian Institute of Public Health. Participation in the HUNT Study is voluntary, and all participants provided written informed consent before participation. 
The Regional Committee on Medical and Health Research Ethics of Norway (REK; 2018/1824) approved this work in July 2021.

\section{Funding}

This work is sponsored by NTNU's Digital Transformation project 'My Medical Digital Twin'.

\bibliographystyle{plainnat}
\bibliography{Bibliography}

\appendix
\clearpage
\pagenumbering{arabic}
\section{The Trøndelag Health Study Protocol}
\label{sec:hunt_protocol}
The HUNT Study protocols are described in detail by \cite{krokstad2013cohort} and \cite{aasvold2021cohort}. Here, we briefly overview the performed data collection relevant to this work. Age and sex were extracted from the Norwegian Population Registry. Height and weight were measured after removing shoes and other heavy clothing. Blood pressure was measured, by trained personnel, in a sitting position after two minutes of rest. Three measurements were taken, one minute apart, of which the mean of the second and third were used to report blood pressure. To assess the current use of blood pressure medication, self-reported questionnaires were used. In HUNT2, the current use captured in "Are you taking medication for high blood pressure?" [Never; Previously; Currently]. In HUNT3, the question was reformulated to "Do you take, or have you taken medication for high blood pressure?" [No; Yes]. Therefore, we combine this question with the answer to "If you are currently taking medicine for high blood pressure, have you felt unwell/ had side effects from this medicine?". We assume only the participant who currently takes medication answered this question. In HUNT4, the use of BP medicine was captured in "Do you currently use any prescription medication for high blood pressure?" [No; Yes]. 

\clearpage
\section{Summary of the HUNT3 Cohort}
\label{sec:summary_hunt3}
The validation data set (HUNT3 cohort) consists of participants with observations of $BP$, $age$, $BMI$ and $sex$  in HUNT3, and $BP$ and missing status in HUNT4. Of $50201$ in HUNT3, $33.3 \%$ drop out prior to HUNT4.
Summary statistics for the HUNT3 cohort are given in \cref{tb:summary_HUNT3}, together with group mean for HUNT3 observations grouped on missing status.

\begin{table}[h]
\centering
\caption{The sample mean and standard deviation
of $BP_F$, $BP_I$, $age$, and $BMI$ and proportion of female/male participants in the HUNT3 cohort are displayed in the third column. The fourth and fifth columns displays sample mean for the present and missing participants in addition to proportions of present/missing participants for the whole cohort and per sex.}
\begin{tabular}{llccc}
\multicolumn{5}{c}{\textbf{Summary of the HUNT3 cohort}} \\ \hline
Variable                            &Unit                       & HUNT3                                     & Present in  HUNT4            & Missing in  HUNT4,    \\ \hline
\multicolumn{1}{l|}{$\text{BP}_4$}  &\multicolumn{1}{l|}{mmHg}  & \multicolumn{1}{c|}{-}                    & 136.48         &   -                  \\
\multicolumn{1}{l|}{$\text{BP}_3$}  &\multicolumn{1}{l|}{mmHg}  & \multicolumn{1}{c|}{133.21 (20.71)}       & 131.48         & 136.67      \\
\multicolumn{1}{l|}{$\text{age}_3$} &\multicolumn{1}{l|}{years} & \multicolumn{1}{c|}{53.08 (16.01)}        & 51.68          & 55.90       \\
\multicolumn{1}{l|}{$\text{BMI}_3$} &\multicolumn{1}{l|}{kg/$m^2$} & \multicolumn{1}{c|}{27.17 (4.41)}      & 27.12          & 27.28       \\
\multicolumn{1}{l|}{$\text{sex}$}   &\multicolumn{1}{l|}{}      & \multicolumn{1}{c|}{}                     &   66.7 \%         & 33.3 \%                  \\
\multicolumn{1}{l|}{\quad female}   &\multicolumn{1}{l|}{0}     & \multicolumn{1}{c|}{54.6 \%}                & 68.7 \%           & 31.3 \%     \\
\multicolumn{1}{l|}{\quad male}     &\multicolumn{1}{l|}{1}     & \multicolumn{1}{c|}{45.4 \%}                & 64.3 \%          & 35.7 \%

\end{tabular}
\label{tb:summary_HUNT3}
\end{table}

\clearpage
\pagenumbering{arabic}
\section{Shared Parameter Model With Additive Effects}
\label{sec:spm_additive}
Based on the work done by \cite{master_Lars} we explore the need for non-linear effects in the SPM introduced in \cref{sec:spm}. We model all continuous variables as additive effects ($f(z)$) through a random walk of order 2 with a sum to zero constraints \citep{gomez2020bayesian}. 
Let $m=1:n$ be the index for the increments and define
    $\Delta^2z_m = z_m - 2z_{m+1}+ z_{m+2} \sim N(0, \sigma^2)$.
The density for $f(\boldsymbol{z})$ is, 
    $f(\boldsymbol{z}|\sigma) \propto \sigma^{-\frac{n-2}{2}} exp\Bigg\{- \frac{1}{2 \sigma} \sum_{m = 1}^{n-2} (\Delta^2z_m)\Bigg\}$.
Further, the 
sum of all random effect components is constrained to be zero. For more information about random walk priors and sum to zero constraints, see \cite{rue2005gaussian}.

The specification of linear predictors of the SPM defined in \cref{sec:spm} becomes,
\begin{align}
\label{eq:spm_additive}
    \eta_{BPi} = \alpha_0 + f_{BP_F}({BP_I}_i) + f_{BP_F}(age_i) + f_{BP_F}({BMI}_i) + \alpha_{sex} sex_i + {\epsilon}_i  \\ \nonumber
    \eta_{mi} = \beta_0 + f_m({BP_I}_i) + f_m(age_i) + f_m({BMI}_i) + \beta_{sex} sex_i 
    + c {\epsilon}_i. \nonumber
\end{align}
All regression parameters $\alpha_0$, $\alpha_{sex}$, $\beta_0$, and  $\beta_{sex}$ are given independent priors $ N(0, {10^3}^2)$. The shared parameters $\epsilon_i$ are assumed to be independent Gaussian $\epsilon_i \sim N(0,\sigma_{\epsilon}^2)$. Both the additive effects $f(.)$ and ${\epsilon}_i$ have hyperparameters (${\sigma_{BP_F}}_{BP_I}$, ${\sigma_{BP_F}}_{age}$, $     {\sigma_{BP_F}}_{BMI}$, ${\sigma_{m}}_{BP_I}$, ${\sigma_{m}}_{age}$, ${\sigma_{m}}_{BMI}$, $ \sigma_{{\epsilon}}$) with independent gamma priors, $\text{Gamma}(1, 5 \cdot 10^5)$. 
The association parameter $c$ is given an informative prior $c \sim N(0, 1^2)$.

\begin{figure}[h]
    \centering
    \includegraphics[width = 0.9\textwidth]{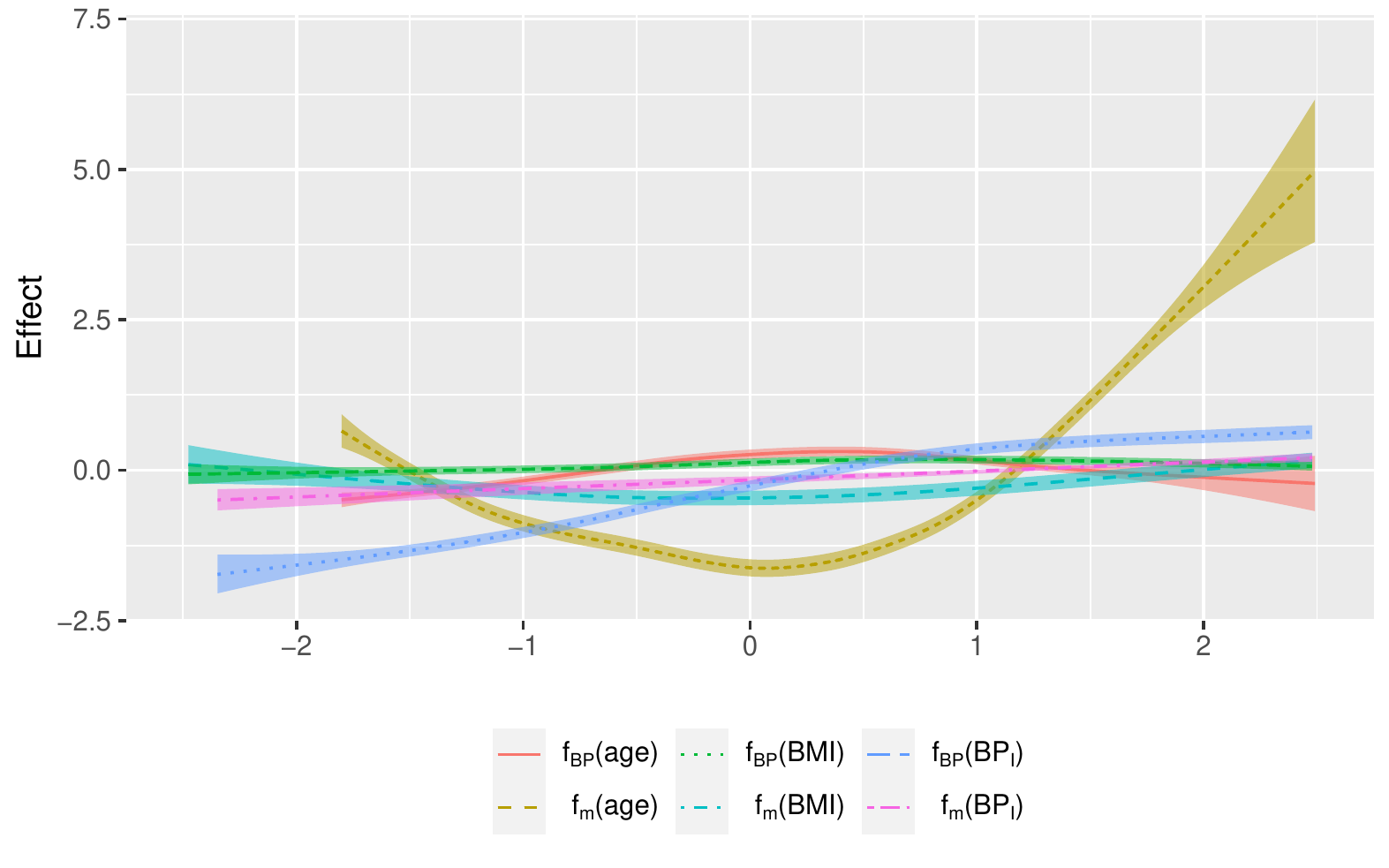}
    \caption{Additive effects of $age$, $BMI$, and $BP_I$ in the SPM for both the $BP$ process and the missing process fitted on 15000 randomly drawn participants from the HUNT2 cohort.}
    \label{fig:additive}
\end{figure}

Modeling all parameters in an additive way is computationally demanding. With 32 CPU cores and 32 GB memory, we were still only able to fit the SPM \cref{eq:spm_additive} with $15 000$ ($\approx 25 \%$) participants.
These participants were drawn randomly. The results are given in  \cref{fig:additive}.
The age effect in the missing process is clearly non-linear. The other continuous variables, although not perfectly linear, are much closer to being linear.

Therefore, we chose to model all variables linearly except for age in the dropout process, which we model as an additive effect. 

\clearpage
\pagenumbering{arabic}

\section{Parameter estimates}
\label{apendix:param_est}
The parameter estimates from the SPM and naive model introduces in \cref{sec:spm} fitted on the HUNT2 cohort, together with their $95\%$ equitailed credible intervals model are presented in \cref{tb:summary_param_est}.

\begin{table}[h!]
\centering
\caption{Summary of all parameters estimated for both the SPM and the naive model fitted on teh HUNT2 cohort. The posterior means and $95 \%$ credible intervals are displayed.}
\begin{tabular}{lcccc}
\hline
\textbf{}                               & \multicolumn{2}{c}{\textbf{SPM}}                       & \multicolumn{2}{l}{\textbf{Naive}} \\
                                        & Posterior mean & \multicolumn{1}{c|}{CI}                  & Posterior mean  & CI               \\ \hline
\multicolumn{1}{l|}{$\alpha_0$}         & 0.275          & \multicolumn{1}{c|}{(0.259, 0.291)}      & 0.134          & (0.123, 0.145)   \\
\multicolumn{1}{l|}{$\alpha_{age}$}   & 0.244          & \multicolumn{1}{c|}{(0.232, 0.255)}      & 0.246          & (0.235, 0.258)    \\
\multicolumn{1}{l|}{$\alpha_{BMI}$}   & 0.073          & \multicolumn{1}{c|}{(0.064, 0.082)}      & 0.071          & (0.062, 0.080)   \\
\multicolumn{1}{l|}{$\alpha_{BP}$}    & 0.599          & \multicolumn{1}{c|}{(0.589, 0.610)}      & 0.578          & (0.567, 0.588)   \\
\multicolumn{1}{l|}{$\alpha_{sex}$}     & 0.041         & \multicolumn{1}{c|}{(0.025, 0.057)}       & -0.022         & (-0.064, 0.019) \\
\multicolumn{1}{l|}{$\beta_0$}          & 0.567          & \multicolumn{1}{c|}{(0.475, 0.666)}      & 0.55           & (0.460, 0.651)                \\
\multicolumn{1}{l|}{$\beta_{BMI}$}    & 0.087          & \multicolumn{1}{c|}{(0.087, 0.106)}      & 0.081          & (0.063, 0.099)                \\
\multicolumn{1}{l|}{$\beta_{BP}$}     & 0.139          & \multicolumn{1}{c|}{(0.139, 0.162)}      & 0.133          & (0.111, 0.155)                \\
\multicolumn{1}{l|}{$\beta_{sex}$}      & 0.292          & \multicolumn{1}{c|}{(0.292, 0.329)}      & 0.275          & (0.240, 0.310)                \\
\multicolumn{1}{l|}{$\sigma_{age}$}   & 1.412          & \multicolumn{1}{c|}{(0.946, 2.130)}      & 1.400            & (0.927, 2.123)               \\
\multicolumn{1}{l|}{$\sigma_{\epsilon}$}& 0.790          & \multicolumn{1}{c|}{(0.783, 0.797)}      & 0.77           & (0.765, 0.776)   \\
\multicolumn{1}{l|}{c}                  & 0.705          & \multicolumn{1}{c|}{(0.645, 0.765)}      & -              & -                \\ \hline
\end{tabular}
\label{tb:summary_param_est}
\end{table}

\clearpage
\pagenumbering{arabic}
\clearpage
\section{Prior Sensitivity Analysis}
\label{sec:prior_sensitivity}

\begin{table}[h]
\centering
\small
\begin{tabular}{l|lllllll}
Model name & $c\_0\_1$  &$c\_0\_10$ &$ c\_0\_100$  & $c\_1\_1$ &$c\_1\_10$ &$c\_1\_100$  &$c\_10\_100$ \\ \hline
Prior for c & N(0,$1^2$) & N(0,$10^2$) & N(0,$100^2$) &    N(1,$1^2$) & N(1,$10^2$) & N(1,$100^2$)  & N(10,$100^2$)      
\end{tabular}
\caption{Priors for association parameter c.}
\label{tb:priors}
\end{table}

As the SPM model specified in \Cref{sec:spm} is rather complex, we perform a sensitivity analysis to evaluate if the model is sensitive to the choice of prior for the association parameter $c$ \eqref{eq:spm}. This parameter is of special interest as it defines the connection between the dropout process and the measurements process, $BP_F$. We fit the model defined in \eqref{eq:spm} with different choices for the prior for $c$ displayed in \Cref{tb:priors}.

\begin{figure}[h]
    \centering
    \includegraphics[width = \textwidth]{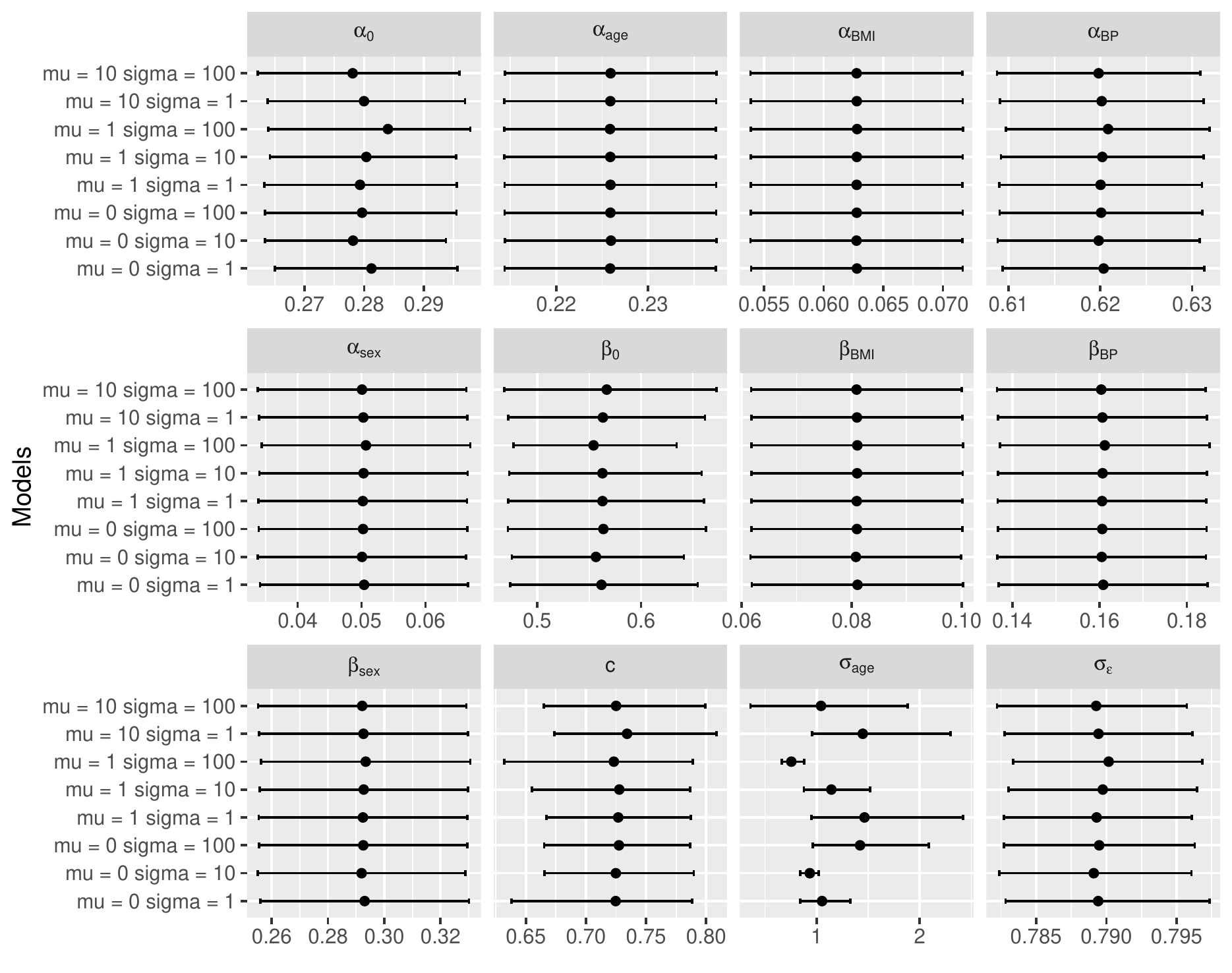}
    \caption{Posterior mean and 95\% credible intervals for all parameters of the SPM with different priors for the association parameter $c$. The model names are constructed so the first digit is the mean and the second the is the standard deviation used in the prior for c.}
    \label{fig:prior_sensitivity}
\end{figure}
\begin{figure}
    \centering
    \includegraphics[width = \textwidth]{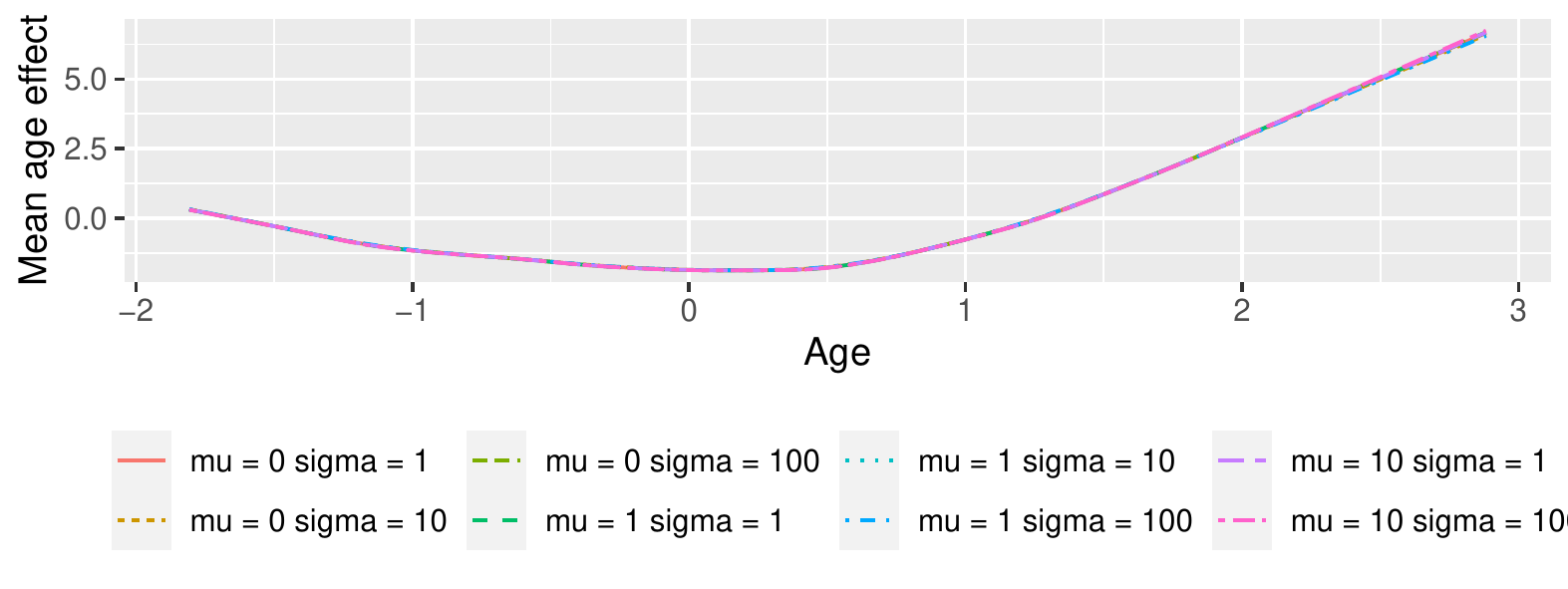}
    \caption{Mean posterior age effect for all models tested in the prior sensitivity study.}
    \label{fig:prior_age_effect}
\end{figure}

\Cref{fig:prior_sensitivity} shows the resulting $95\%$ equi-tailed credible intervals for all parameters. We clearly see that the latent field, $\alpha_0$,  $\alpha_{BP}$,  $\alpha_{age}$, $\alpha_{BMI}$, $\alpha_{sex}$, $\beta_0$, $\beta_{BP}$, $\beta_{BMI}$, $\beta_{sex}$, have almost identical credible intervals. The credible intervals for the hyperparameter $\sigma_{age}$ vary slightly more. However, we see in \cref{fig:prior_age_effect} that the resulting posterior mean of the age effect is practically identical for all the different priors. Hence we conclude that the model is not sensitive to different choices of $c$.

\clearpage
\pagenumbering{arabic}
\FloatBarrier
\section{Supplementary Material for The Simulation Studies on Bias and Coverage}
\label{sec:supplementary_simulation_identifiability}

We have summarized the results from the simulation study performed in \cref{sec:sim_identifiability} exploring the bias and coverage of the SPM
and the naive model. In addition, we have performed a similar study with data MAR. The distribution of the posterior means from the simulation study when the true parameters are the posterior mean estimates of the SPM and the data is MAR can be seen in \cref{fig:simulation_study_MNAR_spm}. 
\cref{tb:sim_mnar_spm_est}, and \cref{tb:sim_mar_spm_est} display the mean posterior mean, bias, and coverage of the parameter estimates for both the SPM and the naive model for simulated data MNAR and MAR, respectively. Further we display the difference in bias for the SPM and naive model $Bias_{SPM} - Bias_{naive}$ also in \cref{tb:sim_mnar_spm_est}, and \cref{tb:sim_mar_spm_est}. 

\begin{figure}
    \centering
        \includegraphics[width =\textwidth]{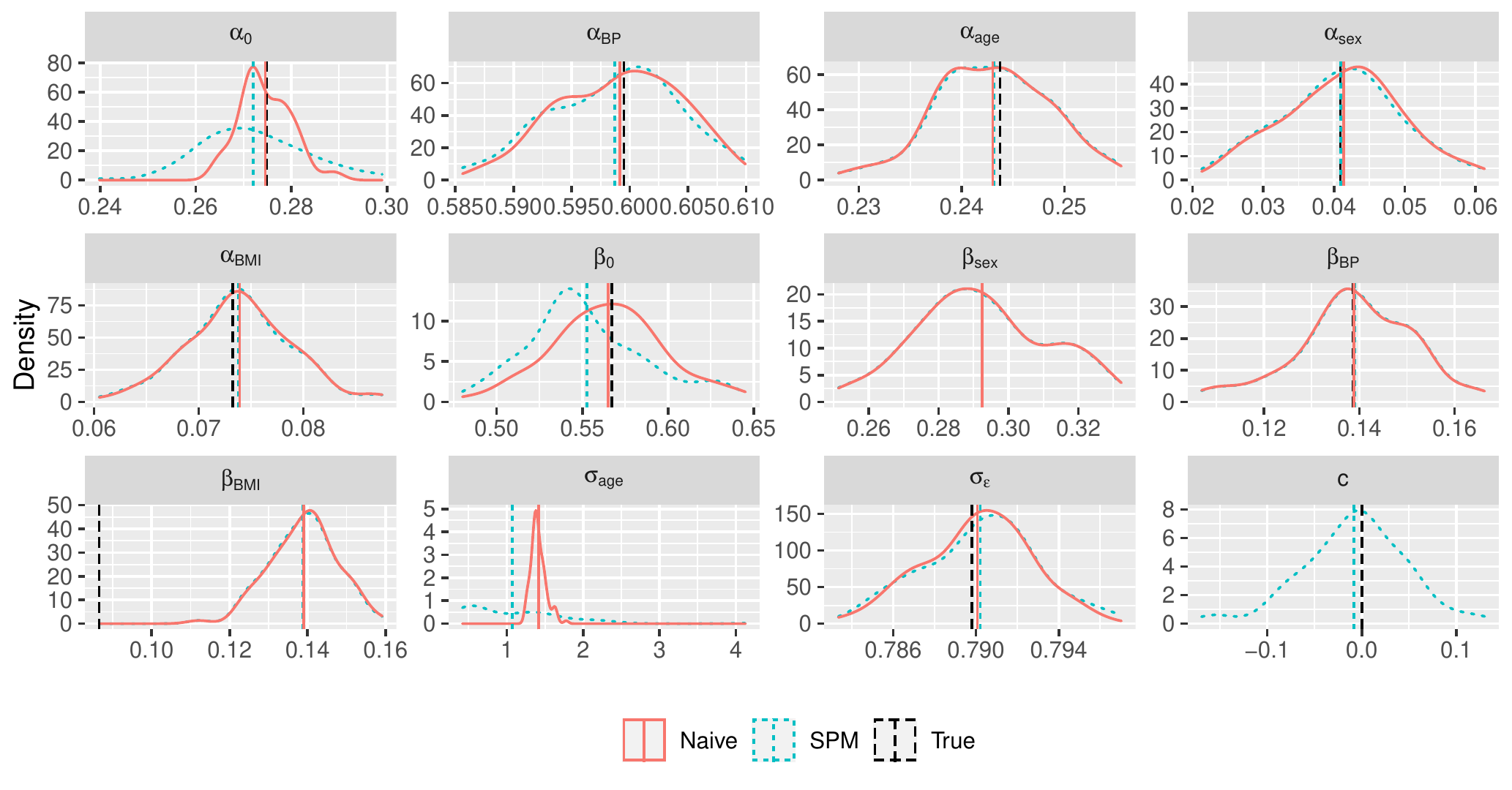}
        \caption{Distribution of posterior mean estimates on simulated data MAR following the SPM. The mean of posterior means for both the SPM and naive model and the true value are indicated by the vertical lines.}
        \label{fig:simulation_study_MAR_spm}
\end{figure}

\begin{table}
\centering
\caption{A summary of the results
where we explore the bias and coverage of the SPM 
and the naive mode
when the true model parameters are known. We display the mean posterior mean, bias of posterior means, and coverage for all parameters. In addition, we display the difference in bias for the SPM and the naive model ($Bias_{SPM} - Bias_{naive}$). The true parameters are the posterior means from the SPM
as given in \cref{tb:summary_param_est} 
and the data is MNAR.}
\begin{tabular}{l|c|ccc|ccc|c}
\textbf{}         & \textbf{}  & \multicolumn{3}{c|}{\textbf{SPM}} & \multicolumn{3}{c|}{\textbf{Naive model}} & \textbf{}                                                     \\\hline
                  & True value & Mean    & Bias      & Coverage   & Mean       & Bias        & Coverage      & \begin{tabular}[c]{@{}l@{}}Difference \\ in bias\end{tabular} \\ \hline
$\alpha_0$         & 0.27       & 0.24    & -3e-02    & 0.07       & 0.13       & -0.144      & 0.00          & 0.112                                                         \\
$\alpha_{BP}$      & 0.60       & 0.60    & -4e-03    & 0.91       & 0.58       & -0.021      & 0.01          & 0.017                                                         \\
$\alpha_{age}$     & 0.24       & 0.24    & -8e-04    & 0.98       & 0.25       & 0.002       & 0.97          & 0.001                                                         \\
$\alpha_{bmi}$     & 0.07       & 0.07    & -1e-03    & 0.95       & 0.07       & -0.006      & 0.76          & 0.004                                                         \\
$\alpha_{sex}$     & 0.04       & 0.04    & -3e-03    & 0.97       & 0.02       & -0.018      & 0.42          & 0.015                                                         \\
$\beta_0$          & 0.57       & 0.56    & -1e-02    & 0.98       & 0.55       & -0.016      & 0.99          & 0.004                                                         \\
$\beta_{BMI}$      & 0.09       & 0.13    & 5e-02     & 0.01       & 0.13       & 0.042       & 0.01          & -0.005                                                        \\
$\beta_{sex}$      & 0.29       & 0.29    & -5e-03    & 0.93       & 0.28       & -0.016      & 0.86          & 0.011                                                         \\
$\beta_{BP}$       & 0.14       & 0.14    & -3e-03    & 0.90       & 0.13       & -0.007      & 0.91          & 0.004                                                         \\
$\sigma_{age}$     & 1.42       & 1.22    & -2e-01    & 0.67       & 1.34       & -0.082      & 1.00          & -0.119                                                        \\
$\sigma_{\epsilon}$ & 0.79       & 0.78    & -8e-03    & 0.32       & 0.77       & -0.020      & 0.00          & 0.012                                                         \\
$c$                 & 0.70       & 0.55    & -2e-01    & 0.00       & NA         & NA          & NA            & NA                                                           
\end{tabular}
\label{tb:sim_mnar_spm_est}
\end{table}

\begin{table}
\centering
\caption{A summary of the results where we explore the bias and coverage of the SPM and the naive model when the true model parameters are known. We display the mean posterior mean, bias of posterior means, and coverage for all parameters. Further, we display the difference in bias for the SPM and the naive model ($Bias_{SPM} - Bias_{naive}$). The true parameters are the posterior means from the SPM 
as given in \cref{tb:summary_param_est} and the data is MAR meaning $c = 0$.}
\begin{tabular}{l|c|ccc|ccc|c}
\textbf{}         & \textbf{}  & \multicolumn{3}{c|}{\textbf{SPM}} & \multicolumn{3}{c|}{\textbf{Naive model}} & \textbf{}                                                     \\ \hline
                  & True value & Mean     & Bias     & Coverage   & Mean       & Bias        & Coverage      & \begin{tabular}[c]{@{}c@{}}Difference\\  in bias\end{tabular} \\\hline
$\alpha_0$         & 0.27       & 0.272    & -3e-03   & 0.8        & 0.27       & -2e-04      & 1.0           & -3e-03                                                        \\
$\alpha_{BP}$      & 0.60       & 0.599    & -8e-04   & 1.0        & 0.60       & -4e-04      & 1.0           & -4e-04                                                        \\
$\alpha_{age}$     & 0.24       & 0.243    & -6e-04   & 1.0        & 0.24       & -7e-04      & 1.0           & 1e-04                                                         \\
$\alpha_{bmi}$     & 0.07       & 0.074    & 6e-04    & 0.9        & 0.07       & 7e-04       & 0.9           & 1e-04                                                         \\
$\alpha_{sex}$     & 0.04       & 0.041    & 7e-05    & 0.9        & 0.04       & 4e-04       & 1.0           & 4e-04                                                         \\
$\beta_0$          & 0.57       & 0.553    & -1e-02   & 1.0        & 0.57       & -2e-03      & 1.0           & -1e-02                                                        \\
$\beta_{BMI}$      & 0.09       & 0.139    & 5e-02    & 0.0        & 0.14       & 5e-02       & 0.0           & 2e-04                                                         \\
$\beta_{sex}$      & 0.29       & 0.292    & -4e-05   & 0.9        & 0.29       & 9e-05       & 0.9           & 4e-05                                                         \\
$\beta_{BP}$       & 0.14       & 0.139    & 3e-04    & 0.9        & 0.14       & 2e-04       & 0.9           & -1e-04                                                        \\
$\sigma_{age}$     & 1.42       & 1.069    & -4e-01   & 0.4        & 1.41       & -6e-03      & 1.0           & -3e-01                                                        \\
$\sigma_{\epsilon}$ & 0.79       & 0.790    & 4e-04    & 1.0        & 0.79       & 3e-04       & 1.0           & -1e-04                                                        \\
$c$                 & 0.00       & -0.008   & -8e-03   & 0.9        & NA         & NA          & NA            & NA                                                           
\end{tabular}
\label{tb:sim_mar_spm_est}
\end{table}


\end{document}